\documentclass[fleqn,usenatbib]{mnras}

\usepackage{newtxtext,newtxmath}
\usepackage[T1]{fontenc}

\DeclareRobustCommand{\VAN}[3]{#2}
\let\VANthebibliography\thebibliography
\def\thebibliography{\DeclareRobustCommand{\VAN}[3]{##3}\VANthebibliography}

\usepackage{graphicx}
\usepackage{amsmath}
\usepackage{siunitx}
\usepackage{float}

\newcommand{\ocen}{$\omega$ Cen}
\newcommand{\solmass}{\(\textup{M}_\odot\)}
\newcommand{\degree}[1]{#1^{\circ}}
\newcommand{\synthdls}{$7.26$ pc}
\newcommand{\synthbhmass}{$19.58$ \solmass}
\newcommand{\synthpm}{1.16 mas/yr}    

\title[Microlensing Survey in Omega Centauri]{A 13 Year Black Hole Microlensing Survey in Omega Centauri}

\author[Zack Freeman et al.]{
Zack Freeman,$^{1}$
Maximilian Häberle,$^{2}$
Anil Seth,$^{1}$\thanks{E-mail: aseth@astro.utah.edu}
John Zaris,$^{3}$
Matthew Whitaker,$^{1}$\newauthor
Carl L. Rodriguez,$^{4}$
Elena Gonz\'{a}lez Prieto,$^{5,6}$
Andrea Bellini$^{7}$,
Callie Clontz$^{10}$,
Sebastian Kamann$^{8}$,\newauthor
Mattia Libralato,$^{9}$
Nadine Neumayer,$^{10}$
Jay Strader,$^{11}$
Do\u{g}a Veske,$^{12,13,14}$
\\
$^{1}$Department of Physics and Astronomy, University of Utah, Salt Lake City, UT 84112, USA\\
$^{2}$European Southern Observatory, Karl-Schwarzschild-Straße 2, 85748 Garching, Germany\\
$^{3}$Department of Physics, University of Colorado, Boulder, CO 80309, USA\\
$^{4}$University of North Carolina at Chapel Hill, USA\\
$^{5}$Department of Physics \& Astronomy, Northwestern University, Evanston, IL 60208, USA\\
$^{6}$Center for Interdisciplinary Exploration \& Research in Astrophysics (CIERA), Northwestern University, Evanston, IL 60201, USA\\
$^{7}$Space Telescope Science Institute, 3700 San Martin Drive, Baltimore, MD 21218, USA\\
$^{8}$Astrophysics Research Institute, Liverpool John Moores University, 146 Brownlow Hill, Liverpool L3 5RF, UK\\
$^{9}$INAF, Osservatorio Astronomico di Padova, Vicolo dell’Osservatorio 5, Padova,I-35122, Italy\\
$^{10}$Max Planck Institute for Astronomy, K\"onigstuhl 17, D-69117 Heidelberg, Germany\\
$^{11}$Department of Physics and Astronomy, Michigan State University, East Lansing, MI 48824, USA\\
$^{12}$Fizik B\"ol\"um\"u, Orta Do\u{g}u Teknik \"Universitesi, \c{C}ankaya, Ankara 06800, T\"urkiye\\
$^{13}$The Research and Application Center for Space and Accelerating Technologies (IVMER), Orta Do\u{g}u Teknik \"Universitesi, \c{C}ankaya, Ankara 06800, T\"urkiye\\
$^{14}$Columbia Astrophysics Laboratory, Columbia University in
the City of New York, New York, NY 10027, USA
}

\date{Accepted XXX. Received YYY; in original form ZZZ}

\pubyear{\the\year{}}

\begin{document}
\label{firstpage}
\pagerange{\pageref{firstpage}--\pageref{lastpage}}
\maketitle

\begin{abstract}
Star clusters are born with massive stars that leave behind a large population of stellar mass black holes. These black holes may account for a large fraction of binary black hole mergers seen in gravitational waves as the dense environment enables retention of black holes and the formation of binaries. However, direct measurements of stellar mass black holes in globular clusters are limited, with just a few found in binary systems thus far. In this paper, we examine the possibility of detecting black holes through microlensing in the Milky Way's most massive star cluster, $\omega$~Centauri. We take advantage of a unique dataset of nearly 200 single-band images taken over 13 years with the Hubble Space Telescope to look for the $\sim$1 year long events expected from microlensing of cluster stars by stellar mass black holes. We develop a novel cost function to fit the lensing signatures that we verify using synthetic events, quantifying our completeness. These synthetic events are based on a recent Monte Carlo simulation of $\omega$~Centauri. We also use these synthetic events to train a neural network to most efficiently distinguish observed lensing events. Application of this algorithm to our observed data, and subsequent visual follow-up, show no convincing candidate black hole lensing events. Based on analysis of the simulation data and our completeness, we find that no events would be expected in the data contrary to the expectations from earlier work. We release the photometric dataset used in this work for future use.
\end{abstract}


%%%%%%%%%%%%%%%%%%%%%%%%%%%%%%%%%%%%%%%%%%%%%%%%%%

%%%%%%%%%%%%%%%%% BODY OF PAPER %%%%%%%%%%%%%%%%%%

\section{Introduction} \label{sec:intro}

At the end of their lifetimes, massive stars undergo supernova, sometimes leaving behind a stellar mass black hole (BH) \citep[e.g.][]{2020FrASS...7...38M}. Hundreds to thousands of BHs are expected to be born in the early stages of globular cluster (GC) formation, but the populations and distribution of these BHs in the present day is not yet well understood  \citep[e.g.][]{Morscher2015,Kremer2018,ArcaSedda2018,Baumgardt_2019}. These (expected) BHs have added significance because they can potentially be the seeds of binary BH mergers detected in gravitational waves \citep[e.g.][]{Rodriguez2016,Mai2026}. In 2016, LIGO-Virgo-Kagra made the first detection of gravitational waves produced from a BH-BH binary merger event \citep{Abbott_2016}. The recent release of the Gravitational-Wave Transient Catalog 5.0  \citep{gwtc5} brings the total number of BH-BH mergers to 390. The mass and spin distributions of this population of gravitational wave events suggests multiple formation pathways, with the most massive BH-BH mergers likely resulting from hierarchical merging in massive star clusters \citep{Antonini2025,gwtc5_pop}.  

Despite this significance, direct detections of stellar mass BHs in GCs remain very limited. There are just a few BHs dynamically detected in binary systems with a visible companion \citep{Giesers2018,Giesers2019,Whitaker2026}, and a handful of other BH candidates detected through accretion from binary companions  \citep[e.g.][]{Maccarone2007,Strader2012,Chomiuk2013,MillerJones2015}. These observations probe only BHs that are in binaries with stellar companions and thus provide only limited evidence on the overall population of BHs. Dynamical and morphological modeling of GCs can provide more indirect evidence of the size of BH populations \citep[e.g.][]{Weatherford2020,Dickson2024}.

The uncertainty on the size of the population of stellar mass BHs in clusters is due to poorly constrained natal kicks and dynamical evolution. BHs can be ejected from a cluster during their formation due to both natal kicks during supernovae and kicks received when a binary companion rapidly loses mass (Blaauw kicks)  \citep[e.g.][]{Popov2025}, if these kicks exceed the cluster escape velocity. This means that clusters with high escape velocities (correlating with larger masses) are likely to retain a larger fraction of BHs. Simulations of dynamical evolution of BHs  \citep[e.g.][]{Morscher2015,Baumgardt2017,Kremer2018,ArcaSedda2018} show that BHs dynamically segregate into the core of the cluster delaying core collapse.  While initially it was thought that the segregation of the black holes would quickly result in their ejection \citep[e.g.][]{Spitzer1969,Kulkarni1993}, it was later realized that interactions between the cluster stars and black hole subsystem slow this process allowing for the retention of black holes in present day clusters \citep[e.g.][]{Morscher2013,Sippel2013}. The BH populations of modern Monte Carlo and N-body simulations depend on assumptions in stellar evolution, as well as the initial properties of the star clusters.

To better understand the distribution of BHs in GCs, the globular cluster \ocen\ is a prime candidate. \ocen\ is the most massive globular cluster observed within the Milky Way, with a present-day mass of \num{3.55e6}\,\solmass \citep{Baumgardt_2018}. This mass is helpful in a few ways: (1) there should be a higher number of BHs initially formed, and (2) the escape velocity is higher, thus likely retaining a larger fraction of the BHs formed. \ocen{} is suspected to have originated from a Nuclear Star Cluster (NSC) \citep[e.g.][]{Freeman1993,Pfeffer2021,Souza2026}, resulting in an even higher escape velocity initially, allowing for more BHs to be retained and hierarchical BH growth. 

Dynamical studies have shown that there is a need for dark mass at the center of \ocen. There has been extensive discussion on whether this suspected dark mass is an intermediate-mass BH (IMBH) \citep{2008ApJ...676.1008N,Noyola_2010,2017MNRAS.464.2174B} or can be explained by other properties, such as a population of stellar mass BHs \citep{2010ApJ...710.1032A,2010ApJ...710.1063V, 2017MNRAS.468.4429Z}. It has been shown that a centrally concentrated cluster of $\mathcal{O}(10^4)$ stellar mass BHs can explain the observed kinematics \citep[e.g.][]{2019MNRAS.482.4713Z, Baumgardt_2019,BanaresHernandez2025}. The latest addition to this debate has been the discovery of various fast-moving stars in the innermost region of \ocen{} \citep{Haeberle_2024a}, which provide strong evidence for an IMBH of at least $8200$~M$_\odot$, but does not exclude a significant mass contribution from stellar mass BHs.  The recent simulations from \citet{prieto_2025} show that the velocity dispersion and surface brightness profile of \ocen{} can be replicated with a simulation that includes a $\sim$50,000~M$_\odot$ IMBH along with $\sim$2500 stellar mass BHs.  But the exact mass of both the IMBH and the number and radial distribution of stellar mass black holes remains poorly constrained.

\citet{Zaris_2020} found that a population of $\sim$10,000 stellar mass BHs could result in detectable self-lensing of stars by stellar mass BHs. Assuming a flat BH density distribution within the core of the cluster and varying the number and mass of BHs they found that significant magnification events of 1.1 (i.e. brightening of 10\%) would occur once every 5--20 years depending on the radial profile of those black holes.  

In addition to being the most massive globular cluster within our Galaxy, \ocen\ has been observed hundreds of times by the \textit{Hubble Space Telescope} (HST) in the past 20 years. This combination of high frequency and long-baseline observations is key to being able to detect a potential gravitational lensing event on the timescale of lensing events that we expect ($\sim1$ year, see Section \ref{subsec:synthetic_events}). In this study, we make use of these data to search for photometric microlensing signatures indicating the presence of stellar-mass BHs in the center of \ocen. 

This paper is organized as follows. In Section \ref{sec:obs_data}, the observational data is described, in addition to detailing cuts made to ensure quality lightcurves. Section \ref{sec:detection_method} describes the star cluster simulations used, as well as the models used to detect lensing signatures and the synthetic events used to tune said models. Then, in Section \ref{sec:synthetic_analysis}, we describe model completeness and our binary classification models. Finally, in Section \ref{sec:bh_pop_disc}, the results of this survey will be discussed.

\section{Observations and Data Reduction} \label{sec:obs_data}
\subsection{Observations}
Due to the high uniform stellar density, the core region of \ocen{} is an ideal target for photometric and astrometric calibrations for high-resolution imaging instruments. For this reason, it has been imaged hundreds of times with various instruments of HST. The highest number of individual exposures has been taken with the Wide Field Camera 3 / Ultraviolet Visual Channel (WFC3/UVIS) and using the filter F606W. These repeat observations were originally intended to monitor the stability of the distortion correction of the instrument (see e.g. \citealt{2015wfc..rept....2K}), but have also been proven to be a unique and valuable scientific dataset. Science results enabled  by these data include deep color-magnitude diagram studies \citep[e.g.][]{bellini_catalog,bellini_populations}, the detection of astrometric binaries (\citealt{2024ApJ...963...60P}, \citealt{Whitaker2026}), and the discovery of fast-moving stars in the innermost arcseconds \citep{Haeberle_2024a}.

In this work, we analyzed a total of 192 individual images taken with WFC3/UVIS F606W. The data have individual exposure times ranging from 4 to 800 seconds and were taken between the year 2009.53 and 2022.49.

\begin{figure}
    \centering
    \includegraphics[width=1\linewidth]{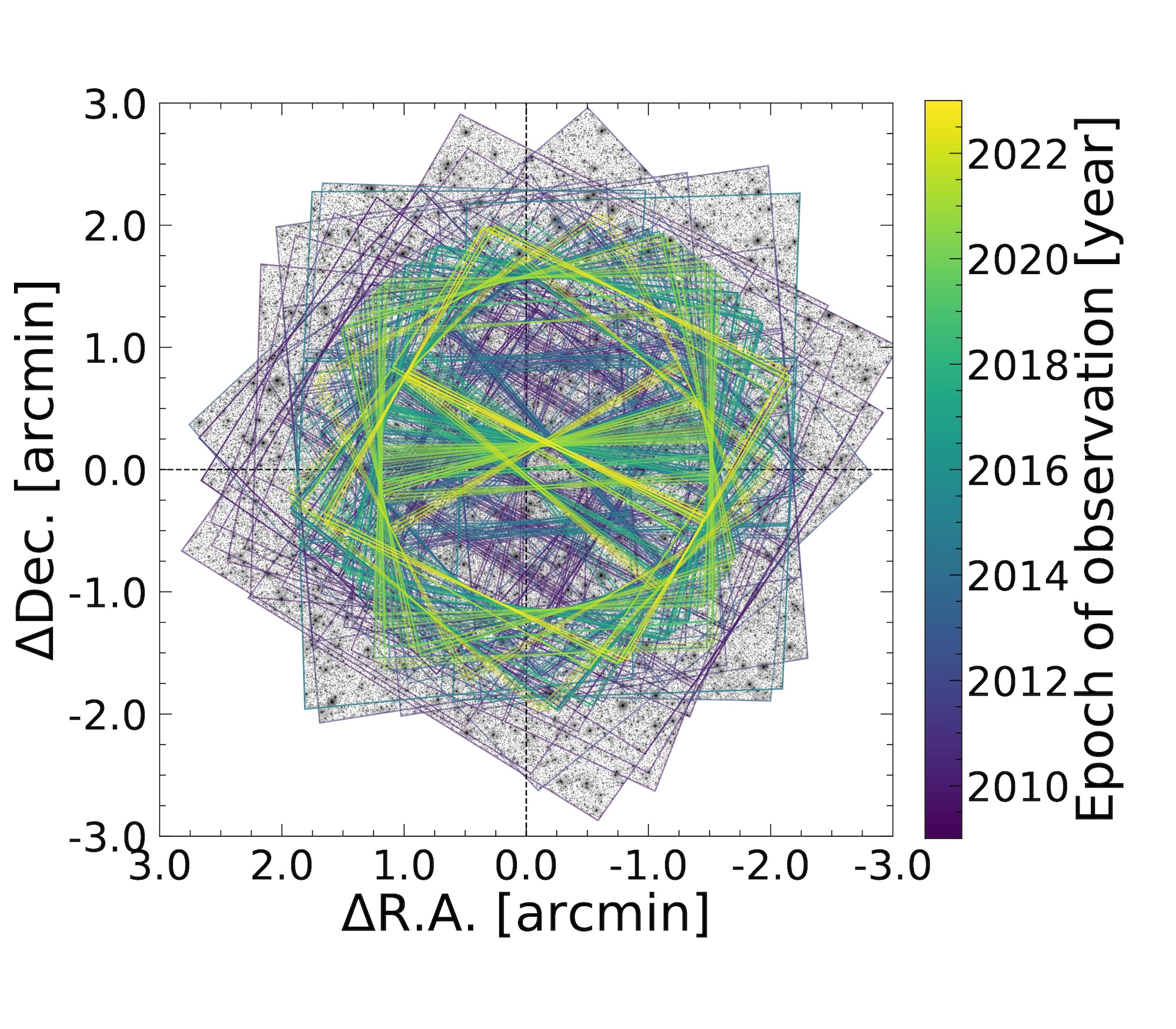}
    \caption{Footprints of the \textit{HST} observations of the core of \ocen{} with the WFC3/UVIS instrument and the F606W filter. The footprints are color-coded by the epoch of the respective observation. In the background, we show a stacked image combining all observations.}
    \label{fig:footprints}
\end{figure}

\subsection{Data Reduction} \label{sec:data_reduction}
The overall data reduction follows the strategy presented in detail in \citet{Haeberle_2024b} (H24b): An initial photometry pass is performed on individual HST images using \texttt{hst1pass} \citep{Anderson2022}, producing source catalogs and optimized empirical point spread function (PSF) models for each image. After registering the individual images on a common astrometric reference frame, a second photometry iteration is performed using the software \texttt{KS2} (see \citealt{bellini_catalog}), which uses the combined images to detect sources (yielding better completeness for faint sources) and uses an iterative PSF photometry scheme, in which neighboring sources are subtracted before each star is measured. 
As the main focus of this study is an accurate time-series using the WFC3/UVIS F606W filter, the exact setup of the \texttt{KS2} reduction is slightly different than in H24b: While H24b split the data into several individual epochs to improve the completeness of fast-moving stars, in this work we use a single combined epoch for all F606W observations. This has the advantage that a consistent set of stars is used for the neighbor subtraction for all individual images, ensuring photometric consistency. We note that this step reduces the completeness for stars that have moved more than 0.5 pixel with respect to their mean position; however, no cluster stars are expected to be this fast (0.5 pixel over 6.5 years corresponds to a proper motion of more than 3\,mas\,yr$^{-1}$, around 4 times the central velocity dispersion of \ocen).
The \texttt{KS2} outputs include photometry using three different methods (see next paragraph) and remapped versions of all input images, which we use to inspect individual measurements.

The \texttt{KS2} measurements used in this work have three photometric methods available for use, which result in potentially varying amounts of noise in the lightcurve. Method One uses the source centroid in the individual image and fits the PSF in a $5 \times 5$ pixel region around this. Method Two determines the best-fit flux using a PSF fit to a $3 \times 3$ pixel region around the source, while Method Three uses only the brightest four pixels weighted by the expected values of the PSF in those pixels. Method One is the most accurate method for brighter isolated sources. However, if the source is not bright enough to produce a distinct peak in the neighbor-subtracted single exposure, then methods two and three may be better suited, as both fix the position of each source based on its position in all the images. See \cite{bellini_catalog} for a detailed description of each method. Due to these differences, sources may have less noise in one photometric method than in another.

\subsection{Ensuring Lightcurves Quality}

\subsubsection{Exposure Times}

To ensure that our lightcurves in the dataset have a similar signal-to-noise ratio and saturation limit across all epochs, we begin by removing individual exposures that have an exposure time less than 35 seconds (8 exposures) or an exposure time greater than 100 seconds (10 exposures). This ensures that data quality is consistent across exposures. This leaves 174 exposures in F606W, most with an exposure time of 40\,s or 60\,s.  

\subsubsection{Estimating Photometric Errors} \label{subsubsec:photometric_errors}

To quantify the significance of potential microlensing events and other photometric outliers, we have to robustly determine the photometric error of each individual measurement. The photometric error model in \citet{Haeberle_2024a} does not include the impact of crowding, and thus we empirically estimate the photometric errors for each star on the individual exposure measurements. $\sigma_G$ is a robust estimate of the standard deviation and is less affected by outliers. It is defined as: $\sigma_G = 0.7413 * (q75 - q25)$, where $q75$ and $q25$ are the third and first quartiles, respectively. The constant $0.7413$ scales the inter quartile range to the standard deviation of a normal distribution. Given the $\sim1$ year period expected for lensing events (Section \ref{subsec:synthetic_events}) -- which would impact a small fraction of the total dataset ($\lesssim10$\% of a lightcurve's data) -- we don't expect true lensing events to have inflated $\sigma_G$ values.  

We determine which exposures contain more noise compared to others by calculating the normalized residuals in each star $i$:

\begin{equation}\label{eq:norm_residual}
    N_{i,j} = \frac{m_{i, j} - \mu_i}{\sigma_{G, i}}
\end{equation}

Where $m_{i,j}$ is the magnitude of the star $i$ in exposure $j$, $\sigma_{G, i}$ is the $\sigma_G$ of star $i$, and $\mu_i$ is the mean magnitude of star $i$. We then calculate the mean noise in each exposure:

\begin{equation}\label{eq:epoch_noise}
    \overline{N_j} = \frac{\sum_{i}^{M_j} |N_{i,j}|}{M_j}
\end{equation}

Here, $M_j$ is the number of stars observed in exposure $j$. We calculate the average and standard deviation noise across the exposures, removing all exposures with noise greater than a $3\sigma$ threshold. This process is performed iteratively, removing a total of four exposures, leaving us with a maximum of 170 exposures.

\begin{figure}
    \centering
    \includegraphics[width=\columnwidth]{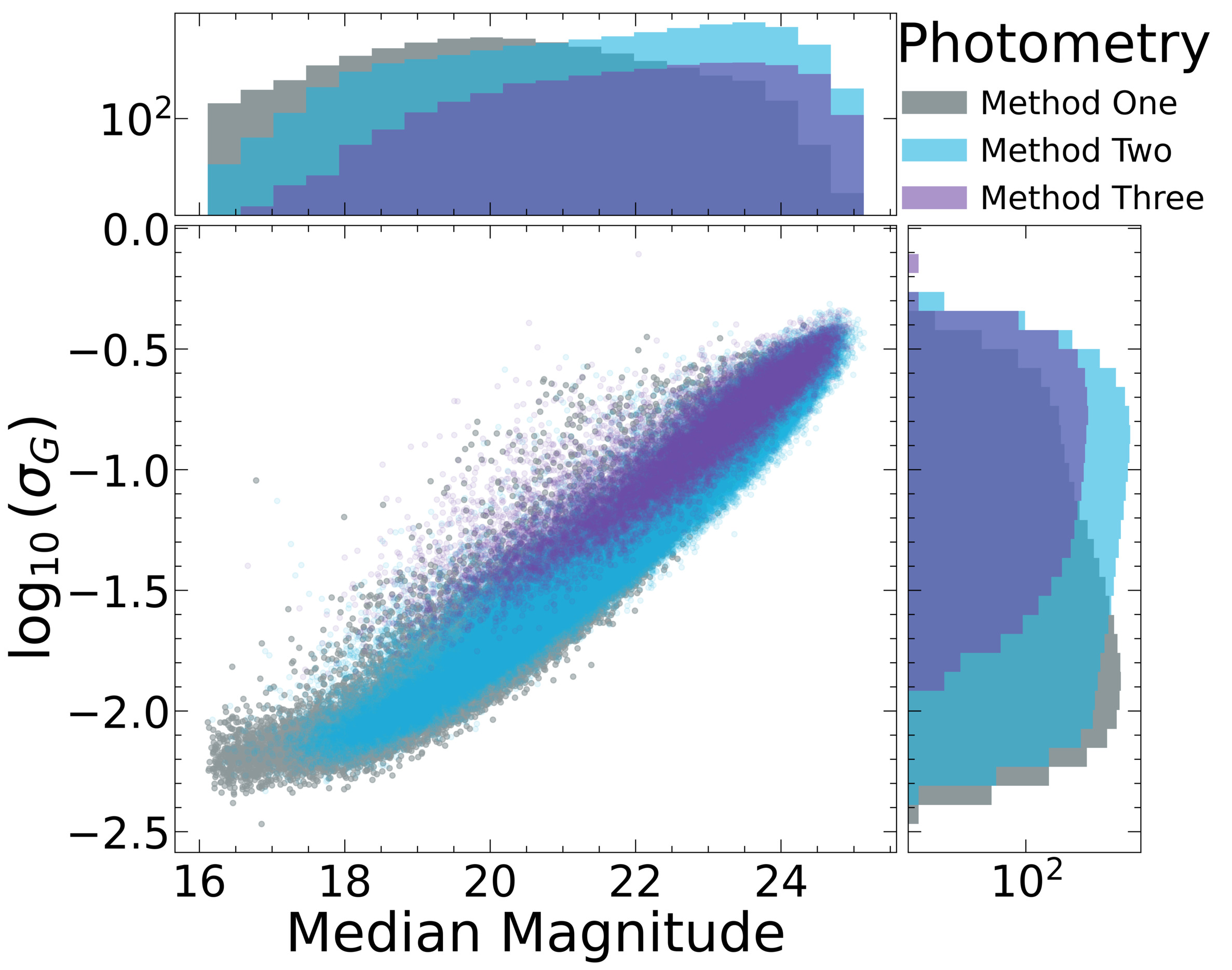}
    \caption{$\sigma_G$ as a function of the median F606W magnitude. Each lightcurve uses the photometric method with the smallest $\sigma_G$ value. In the top histogram, it can be seen that photometry method one dominates at bright magnitudes while method two dominates at dim magnitudes.}
    \label{fig:sigmaG}
\end{figure}

To minimize the noise in the dataset, for each individual star we use the photometric method with the smallest $\sigma_G$ value.  This results in significantly lower $\sigma_G$ for many stars than if we just use instrumental magnitude cut-offs as suggested in \citet{bellini_catalog}.  We can see the best method for each star as a function of magnitude and $\sigma_G$ in Fig.~\ref{fig:sigmaG}; as expected, Method One is typically the best measurement at bright magnitudes, with Methods Two and Three most commonly selected at fainter magnitudes.

\subsubsection{Diffraction Spikes}\label{subsubsec:diff_spikes}

The most prominent photometric contaminants are diffraction spikes. Due to HST's four secondary-mirror support struts, we observe four spikes around bright stars. For each exposure, we compute the spike directions in the image/detector plane directly from the roll angle of HST (which varies between individual exposures): the four spikes are oriented at $0\degree{}$, $90\degree{}$, $180\degree{}$, and $270\degree{}$ relative to the roll angle, and we mask a narrow region along each of these directions centered on any bright star.

\begin{figure}
    \centering
    \includegraphics[width=1\linewidth]{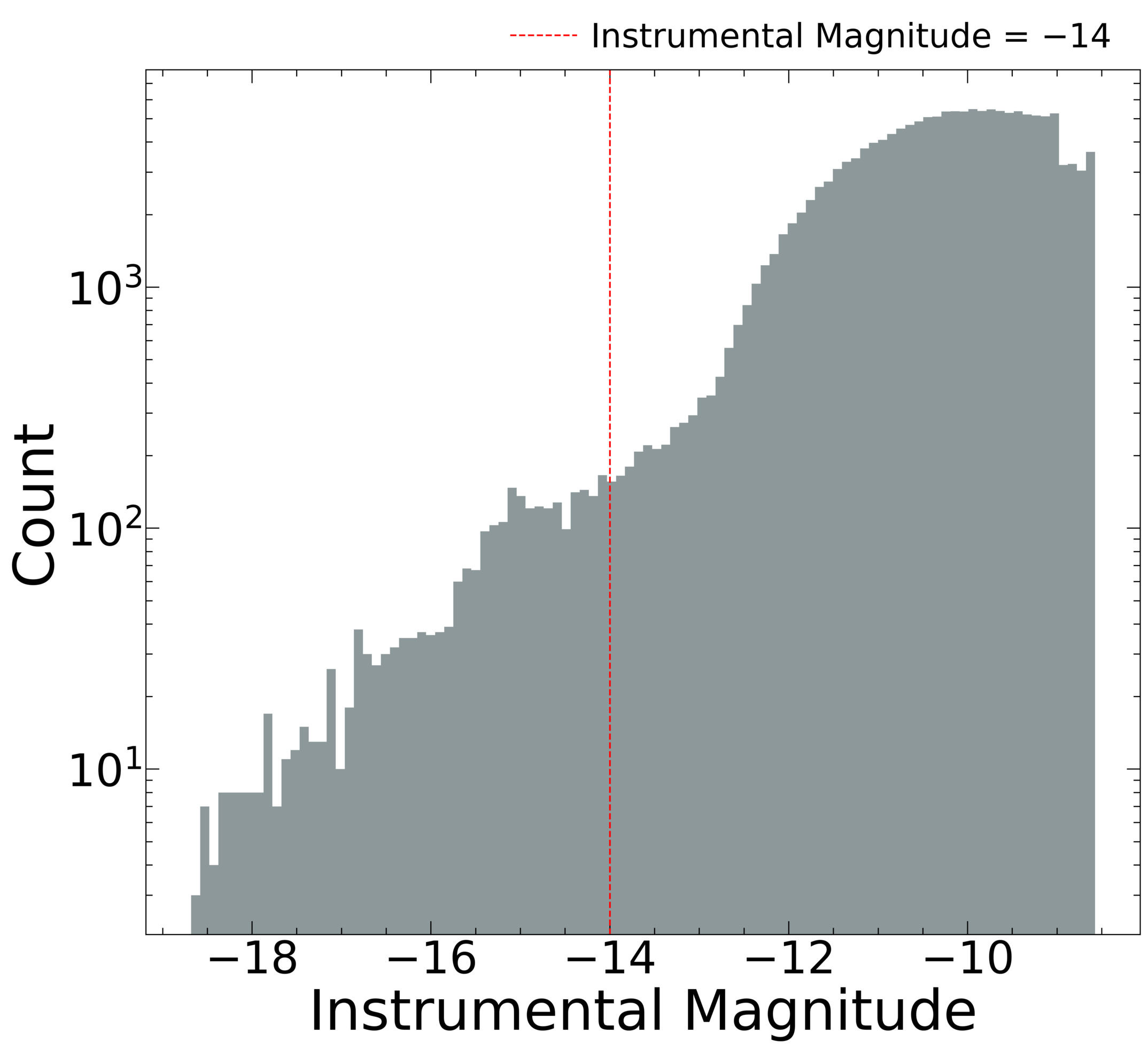}
    \caption{The figure shows the instrumental magnitudes of 157,436 sources. All sources with an instrumental magnitude brighter than $-14$ are considered sources with diffraction spikes (2,584 sources).}
    \label{fig:instr_mag}
\end{figure}

These diffraction spikes extend to large radii for the brightest stars, resulting in them overlapping with nearby stars and causing a temporary increase in their observed brightness that can mimic the lensing signatures we are interested in. Through visual inspection, we found that any star with an instrumental magnitude brighter than $-14$ has visible diffraction spikes. There are $2,584$ stars brighter than this limit that potentially affect surrounding stars (Fig.~\ref{fig:instr_mag}); note that we include stars from the oMEGACat catalog beyond the edge of the observations considered by assuming their F606W magnitude is approximately the same as their F625W magnitude \citep[which are available over a larger field;][]{Haeberle_2024b}.

For each of these bright stars, we discard photometric measurements from stars along each of the four diffraction spikes. We found through visual inspection that the brightest stars have diffraction spikes that reach up to roughly 100 pixels in radius and widths of 10 pixels, while the dimmest stars above the instrumental magnitude threshold of $-14$ have diffraction spikes that reach up to roughly 50 pixels in radius and widths of 10 pixels. Therefore, we used masks of varying radius values dependent upon the instrumental magnitude of the star to ensure high-quality lightcurves while retaining as much data as possible; these radii are given in Table~\ref{table:diff_spike_radii}. However, because of the difficulties of determining the precise radius at which a diffraction spike ends, we applied two masks: the \textit{primary mask}, where observations that fall within this mask are removed from the dataset, and the \textit{extended mask}, where observations that fall within this mask are flagged in the dataset but are not removed. We make one exception to this, where if any observation that falls within the extended mask is brighter than the median magnitude of its star by a difference more than $2\sigma_G$, then the observation is removed instead of flagged. The extended masks extend to radii twice as large as the corresponding primary mask. The width is constant across all masks at 10 pixels.

\begin{table}
	\centering
	\caption{The radius to which a diffraction spike mask extends given an instrumental magnitude. Observations that fall within the primary masks are removed from the dataset, while observations that fall within the extended masks are flagged in the dataset but are not removed, with the exception of bright outliers more than $2\sigma_G$ away from the median magnitude of its star. All masks use a width of 10 pixels. Stars with instrumental magnitudes between -13 and -14 use only an extended mask.}
	\label{table:diff_spike_radii}
	\begin{tabular}{ccc}
		\hline
		Instrumental Magnitude & Primary Radius & Extended Radius\\
		\hline
		$<-13$ & NA & 40\\
		$<-14$ & 60 & 120\\
		$<-16$ & 80 & 160\\
        $<-17$ & 100 & 200\\
		$<-18$ & 120 & 240\\
		\hline
	\end{tabular}
\end{table}

\begin{figure}
    \centering
    \includegraphics[width=1\linewidth]{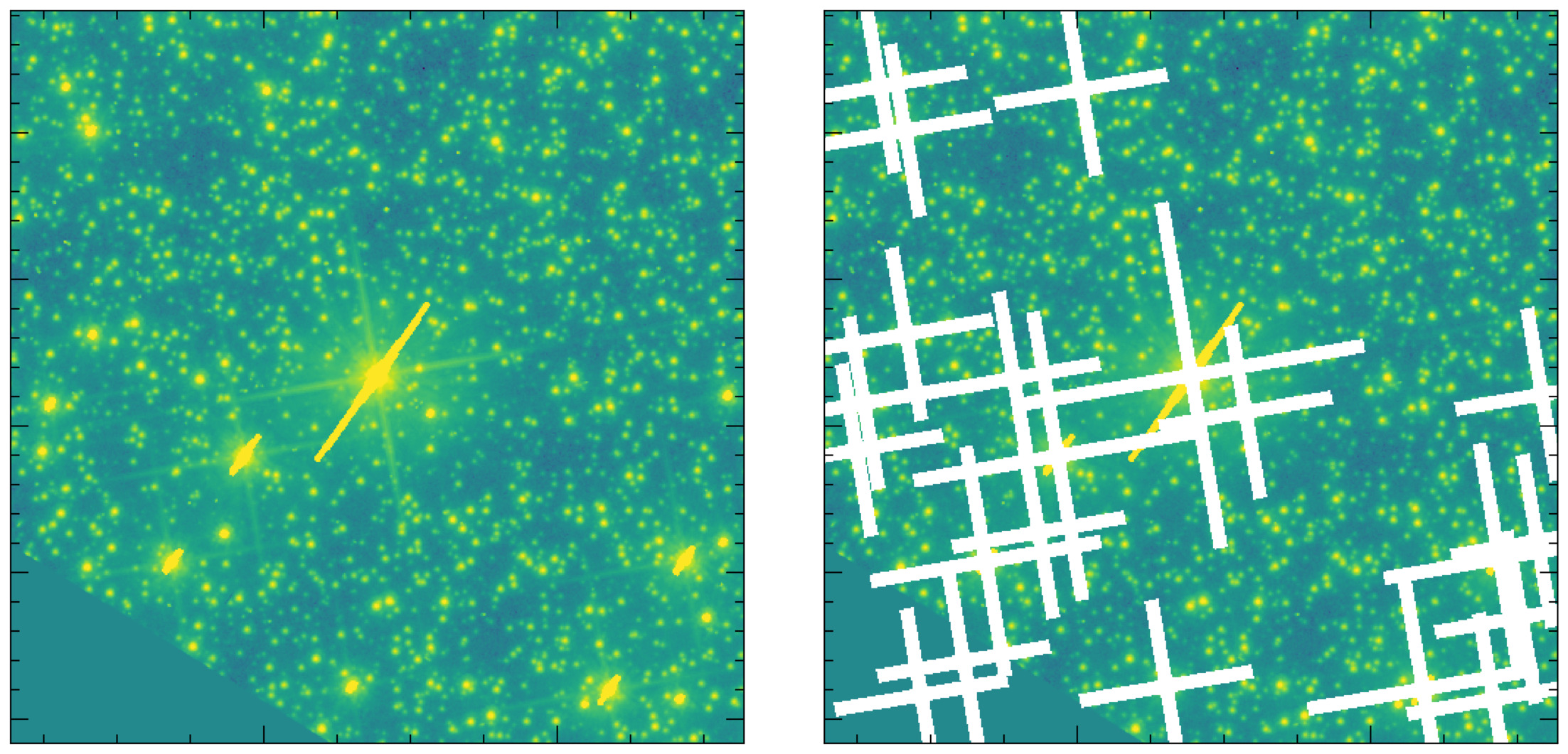}
    \caption{In the center of each panel, a source with an instrumental magnitude of $-18.5$ is shown. In the left panel, the diffraction spikes of the source in the center can be seen extending out to nearly 100 pixels, along with various other dimmer sources with smaller diffraction spikes. In the right panel, the primary mask has been applied to all sources, using the respective radius as seen in Table~\ref{table:diff_spike_radii}. We note that the yellow saturated pixels along the detector columns already are flagged by within the photometric software. We note the coordinate system used here are pixel coordinates aligned to the RA (x-axis) and Dec (y-axis) directions; the same coordinates used in the \citet{Haeberle_2024b} catalog.}
    \label{fig:diff_spike_vis}
\end{figure}

Fig.~\ref{fig:diff_spike_vis} shows an example of a bright star and its diffraction spike mask.  Applying these bright star masks (i.e., the \textit{primary mask}) to each of our exposures ends up removing $12.25\%$ of the individual photometric measurements while $16.51\%$ of the data is flagged by the extended diffraction spike mask. This greatly cleans up our photometry, removing groups of photometric outliers taken at similar roll angles created by the presence of diffraction spikes.  

\subsubsection{Hot Pixels}

We identify hot pixels by looking for an enhancement of outliers at a single detector coordinate (often from measurements of multiple stars).  We derive outliers based on their normalized residuals (equation~(\ref{eq:norm_residual})).  We define outliers as individual measurements with $|N_{i,j}| > 2$. Assuming Gaussian noise on each of our $\sim$200,000 lightcurves, we expect $\sim$2\% of the individual WFC3/UVIS pixels to have a bright outlier measurement (i.e., $N_{i,j} < -2$); thus, the presence of two bright outlier sources centered in a single pixel is a strong indication of a possible hot pixel. We therefore mask individual measurements located in pixels where there are two or more bright outlier sources. The pixels most affected had 10 bright outliers. We find a total of 19,154 hot pixels, resulting in a flagging of 0.24\% of our individual measurements.

\subsubsection{Bright Outliers}

We remove exposures that contain an excess fraction of bright outliers ($2\sigma$ greater than the median fraction), where we again classify bright outliers as observations that are $2\sigma_G$ brighter than the median magnitude of the lightcurve it is a member of. When the bright outlier fraction is visualized against the epoch of exposure, we see roughly a dozen exposures with an unusually large bright outlier fraction. Upon further inspection, we found that this excess bright outlier fraction is correlated with an increase of the PSF median Full Width Half Max (FWHM), possibly related to increased telescope jitter due to gyroscope performance issues  \citep{2018wfc..rept....7A} or focus variations. We remove 15 exposures that have excess bright outlier fractions, leaving us with 155 exposures in the dataset. 

\subsubsection{Final Dataset}\label{subsubsec:final_dataset}

\begin{figure}
    \centering
    \includegraphics[width=1\linewidth]{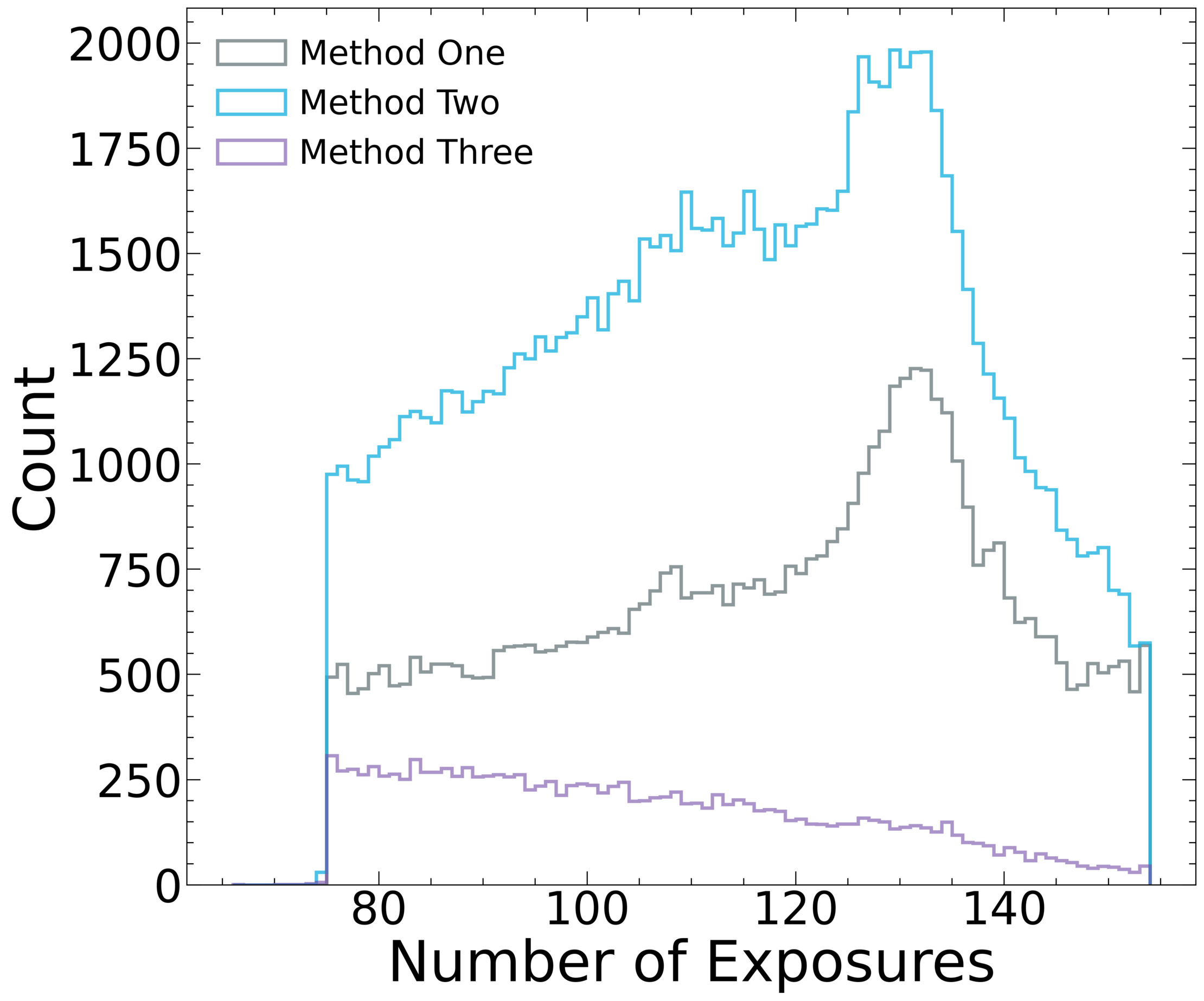}
    \caption{The distribution of the number of exposures in each lightcurve across different photometric methods.}
    \label{fig:obs_dist}
\end{figure}

With the above cuts applied, we enforce a lower limit of 75 exposures in each lightcurve (as measured in photometry Method One). This leaves us with a total of 172,046 lightcurves across 155 exposures. The distribution of the number of exposures across different photometric methods can be seen in Fig.~\ref{fig:obs_dist}. The peak in the distribution in photometric methods one and two is due to the number of available exposures; photometric method three lacks a similar peak since typically this method is used for the faintest stars, which are often not detected in every epoch.  

\section{Detecting Lensing Signatures}\label{sec:detection_method}

The expected microlensing signature is well described using a simple set of equations derived by \cite{Einstein_1936} and taken here from \cite{schneider_2006}.  

\begin{equation} \label{eq:y_scaling}
    y = \frac{\beta}{\theta_E}
\end{equation}
    
\begin{equation} \label{eq:magnification}
    \mu = \frac{y^2 + 2}{y \sqrt{y^2 + 4}} 
\end{equation}

\begin{equation} \label{eq:einstein_angle}
    \theta_E = \sqrt{\frac{4GM}{c^2} \frac{D_{LS}}{D_{S} D_{L}}} 
\end{equation}

\begin{equation} \label{eq:lens_eq}
    \beta = \theta - \theta_E^2 \frac{\theta}{|\theta|^2}
\end{equation}

The magnification, $\mu$ (equation~(\ref{eq:y_scaling}) \& (\ref{eq:magnification})), of a microlensing event can be calculated using the Einstein angle, $\theta_E$ (equation~(\ref{eq:einstein_angle})), and the lens equation (equation~(\ref{eq:lens_eq})). In these equations $G$ is the gravitational constant, $M$ is the mass of the lens, $c$ is the speed of light, $D_{LS}$ is the distance between the lens and the source, $D_S$ is the distance to the source, $D_L$ is the distance to the lens, $\theta$ is the angular position of the deflected light ray in the lens plane, and $\beta$ is the true angular separation of the source and lens from the observer's perspective (if no lensing occurred). For the case of self-lensing within \ocen{} \citep[see also][]{Zaris_2020}, the source and lens distance are both approximately the same; we assume a distance of 5.43 kpc \citep{Baumgardt_2021}. In this case $\theta_E$ depends primarily on the $D_{LS}$ and the lens mass, with larger lens-source distances and lens masses leading to a large $\theta_E$.  

We use these relations both to identify potential lensing events in our data and to create synthetic events; all these steps are described in this section. We use star cluster simulations (Section \ref{subsec:cluster_simulations}) to determine the distributions of microlensing parameters, then use these distributions to create synthetic microlensing events that we inject into our observed data (see Section~\ref{subsec:synthetic_events}) to create realistic synthetic lightcurves. By creating synthetic lightcurves, we can test and optimize our fitting method (Section \ref{subsec:fitting}) and understand what lensing events we can recover (Section \ref{subsec:completeness}).

\subsection{Star Cluster Simulations}\label{subsec:cluster_simulations}

We create synthetic lensing events by extracting typical microlensing parameters from the $N$-body simulations of \ocen{} presented in \citet{prieto_2025}.  These simulations were computed with the Cluster Monte Carlo code \citep{Rodriguez_2022}. Details of the initial conditions are provided in Section 2.1 of \citet{prieto_2025}. The cluster simulations include $1.1 \times 10^{7} $ particles, a cluster mass of roughly $5.5 \times 10^6 M_{\odot}$, and a $500 M_{\odot}$ IMBH seed, which grows over time. The $12$ Gyr snapshot matches the stellar density profile and velocity dispersion profile of the observations, and contains 2,459 stellar mass black holes. The individual masses of these black holes are derived initially from the {\tt COSMIC} code \citep{Breivik2020}, and then the population of black holes are dynamically evolved over 12 Gyr.

\begin{figure*}
    \centering
    \begin{minipage}{1\linewidth}
    \makebox[.5\linewidth]{\includegraphics[width=.5\linewidth]{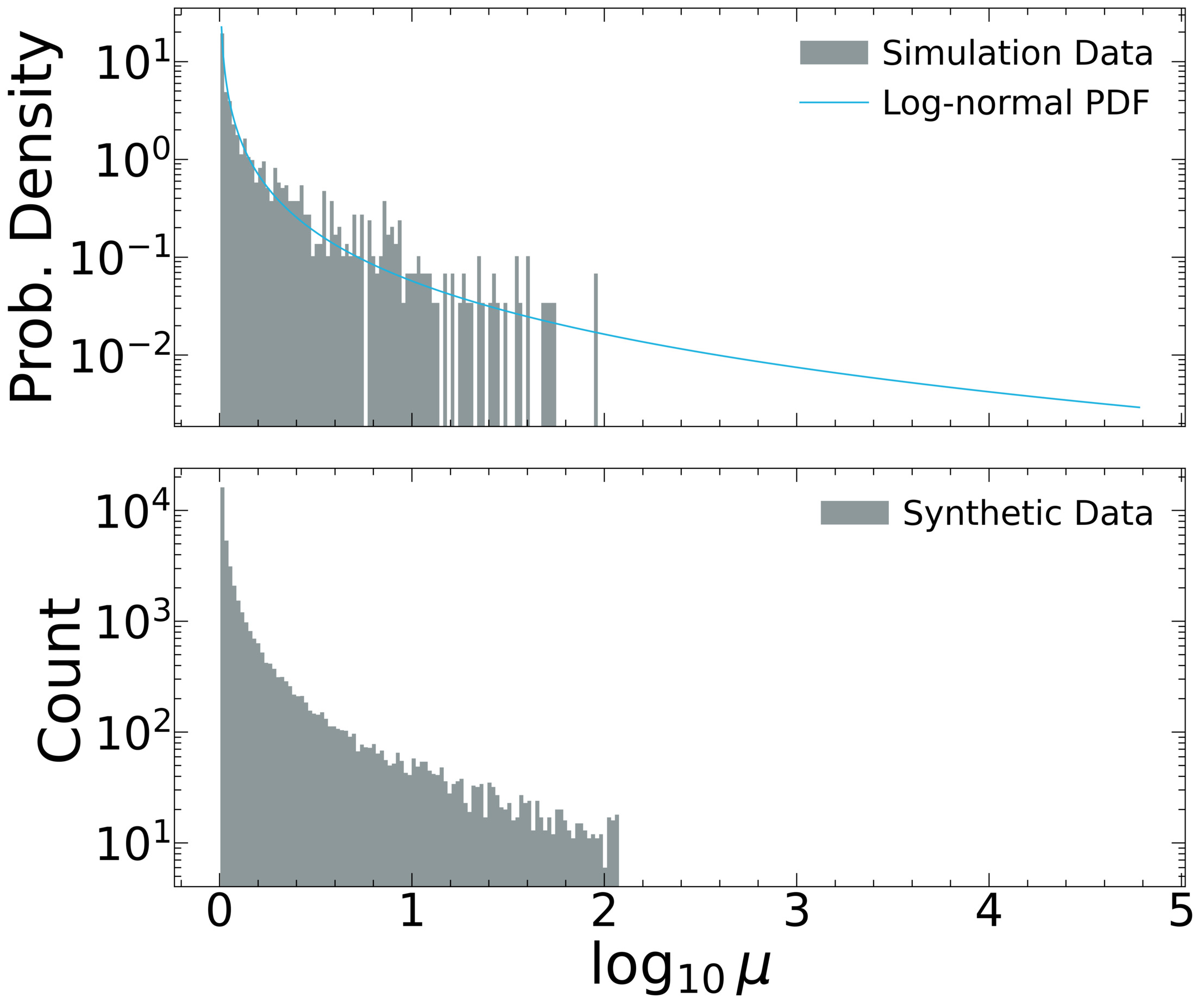}}%
    \makebox[.5\linewidth]{\includegraphics[width=.5\linewidth]{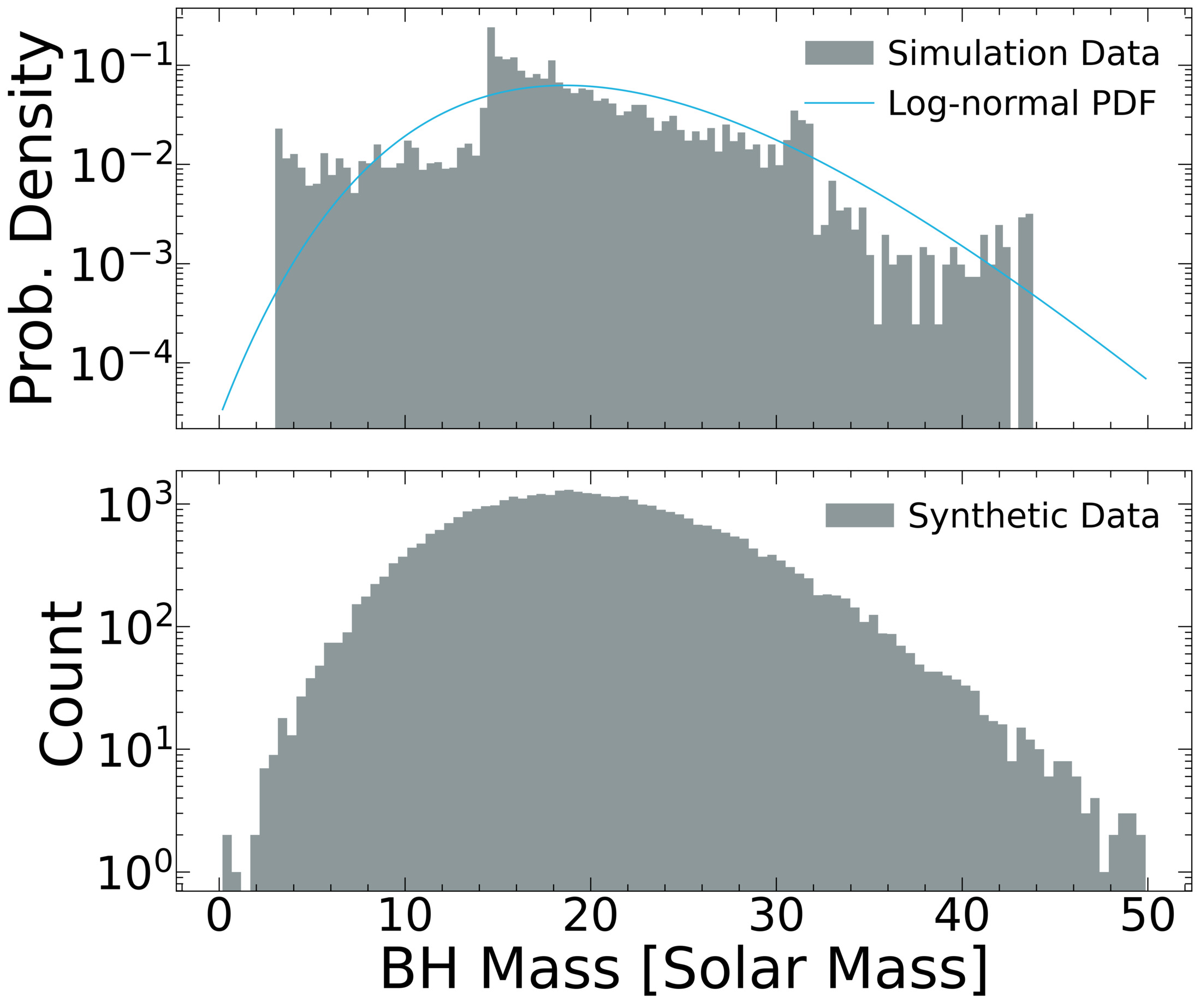}}
    
    \medskip
    
    \makebox[.5\linewidth]{\includegraphics[width=.5\linewidth]{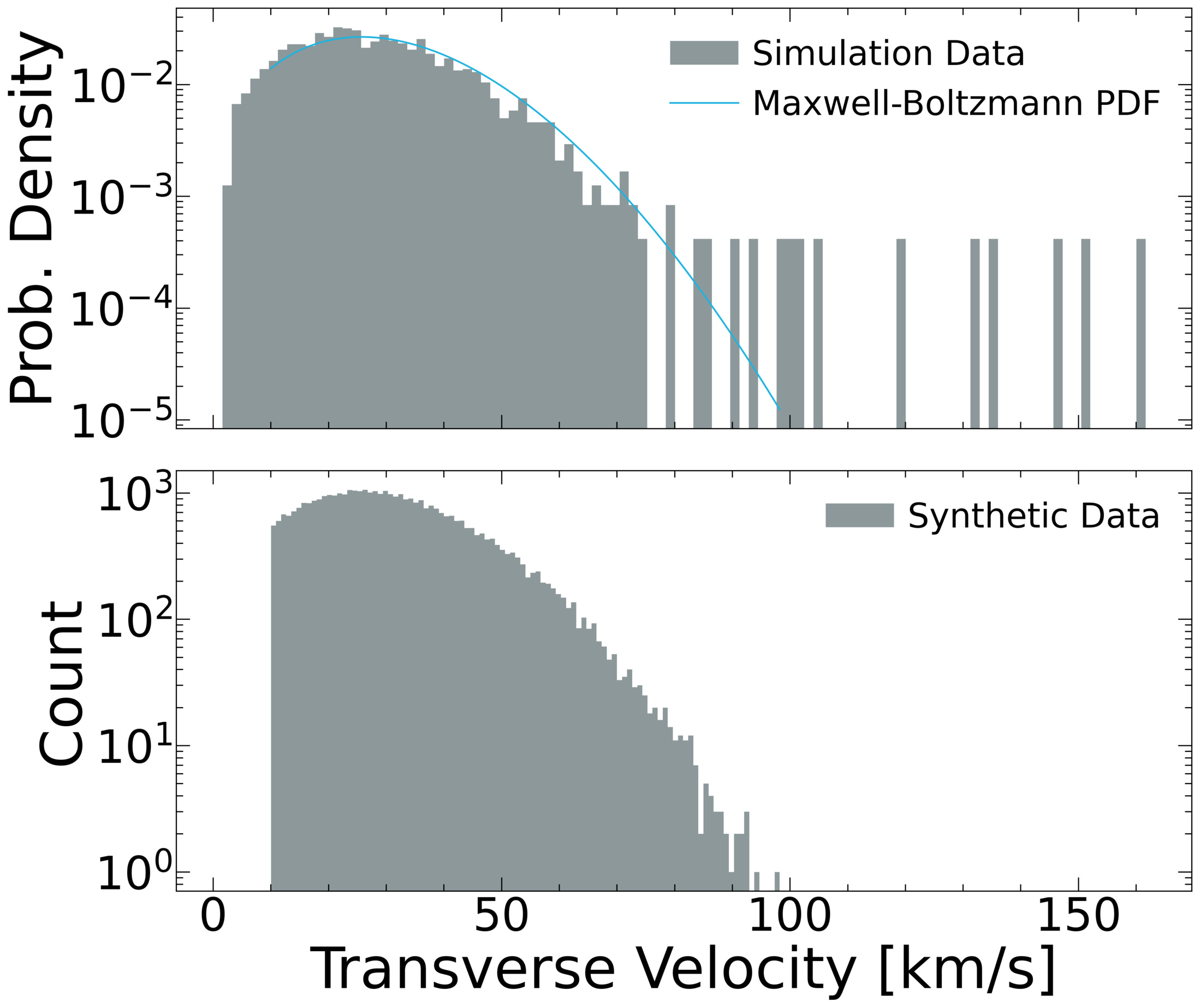}}%
    \makebox[.5\linewidth]{\includegraphics[width=.5\linewidth]{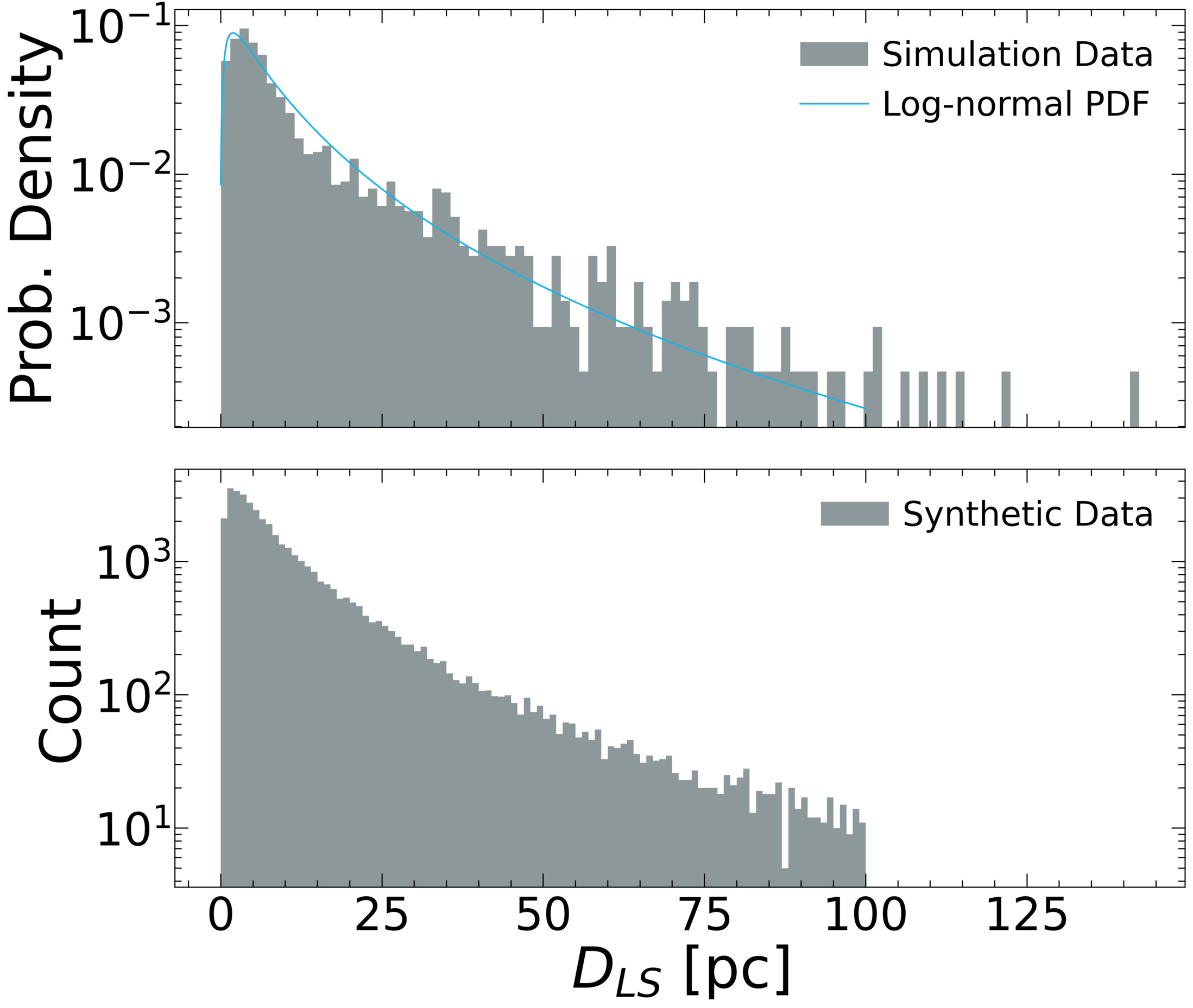}}

\end{minipage}
  
  \caption{In the top panels of this figure, simulation data that reached a maximum magnification greater than $1.01$ is shown with the best-fit PDF. Note that the magnification is shown as $\log_{10}{\mu}$ rather than $\ln{\mu}$ for visualization purposes. In the bottom panels, the distributions of the injected parameters for the 40,000 synthetic lightcurves can be seen. We enforce a minimum transverse velocity of 10 km/s and a maximum $D_{LS}$ of 100 pc to reduce the number of long duration synthetic microlensing events.}
  \label{fig:parameter_distributions}
\end{figure*}

Due to computational limits, we only simulated $10,000$ synthetic lensing events.  Of these 10,000 simulated events, only 1,496 simulations have $\mu \geq 1.01$. Therefore, rather than using individual lensing events derived from the simulations for our synthetic lightcurves, we instead model the \textit{probability density functions} (PDFs) in each of the four parameters that affect a microlensing event ($\ln{\mu}$, where $\mu$ is the maximum magnification, $M$, $D_{LS}$, and $v_T$, the relative transverse velocity) from the 1,496 simulated lensing events with $\mu \geq 1.01$ (Fig.~\ref{fig:parameter_distributions}). Note that we show $\log_{10}\mu$ in Fig.~\ref{fig:parameter_distributions} rather than the natural log for visualization purposes. An explanation of the method used for simulating synthetic lensing events can be found in Appendix Section \ref{app:bh_sampling}.

The $\ln{\mu}$, $M$, and $D_{LS}$ were each fit by log-normal distributions, while $v_T$ was fit by a Maxwell-Boltzmann distribution. The parameters that describe each of the fits can be found in Table~\ref{table:dist_params}.

\begin{table}
	\centering
	\caption{Fitted distributions for each of the four simulation parameters, with mean and standard deviation derived from SciPy \citep{scipy_2020} fits.}
	\label{table:dist_params}
	\begin{tabular}{cccc}
		\hline
		Parameter & Distribution & Mean & Std. Dev.\\
		\hline
		$\ln{\mu}$       & Log-normal        & 0.831   &  7.897\\
        $M$              & Log-normal        & 20.097  &  6.573\\
        $D_{LS}$         & Log-normal        & 14.235  &  23.518\\
        $v_T$            & Maxwell-Boltzmann & 29.724  &  14.826\\
		\hline
	\end{tabular}
\end{table}

With these distributions, we are able to produce random variates, allowing us to produce any number of unique and realistic synthetic lightcurves (discussed in Section \ref{subsec:synthetic_events}).

\subsection{Synthetic Event Simulations}\label{subsec:synthetic_events}

\begin{figure}
    \centering
    \includegraphics[width=1\linewidth]{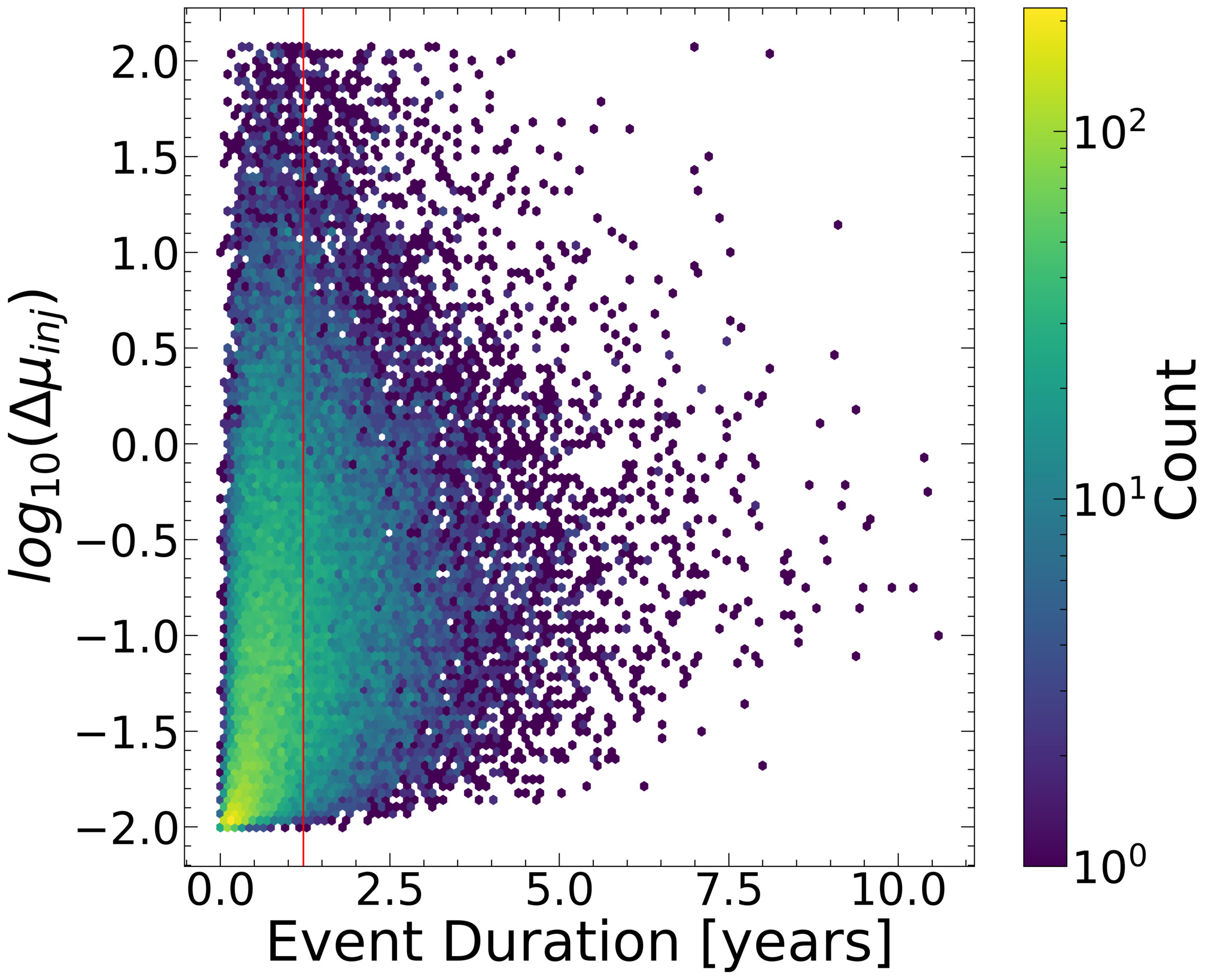}
    \caption{A histogram of the event duration in years vs. the maximum magnification of each synthetic microlensing event. The red line indicates the mean event duration of 1.22 years. The standard deviation of event durations is 1.06 years.}
    \label{fig:mag_vs_duration}
\end{figure}

\begin{figure}
    \centering
    \includegraphics[width=1\linewidth]{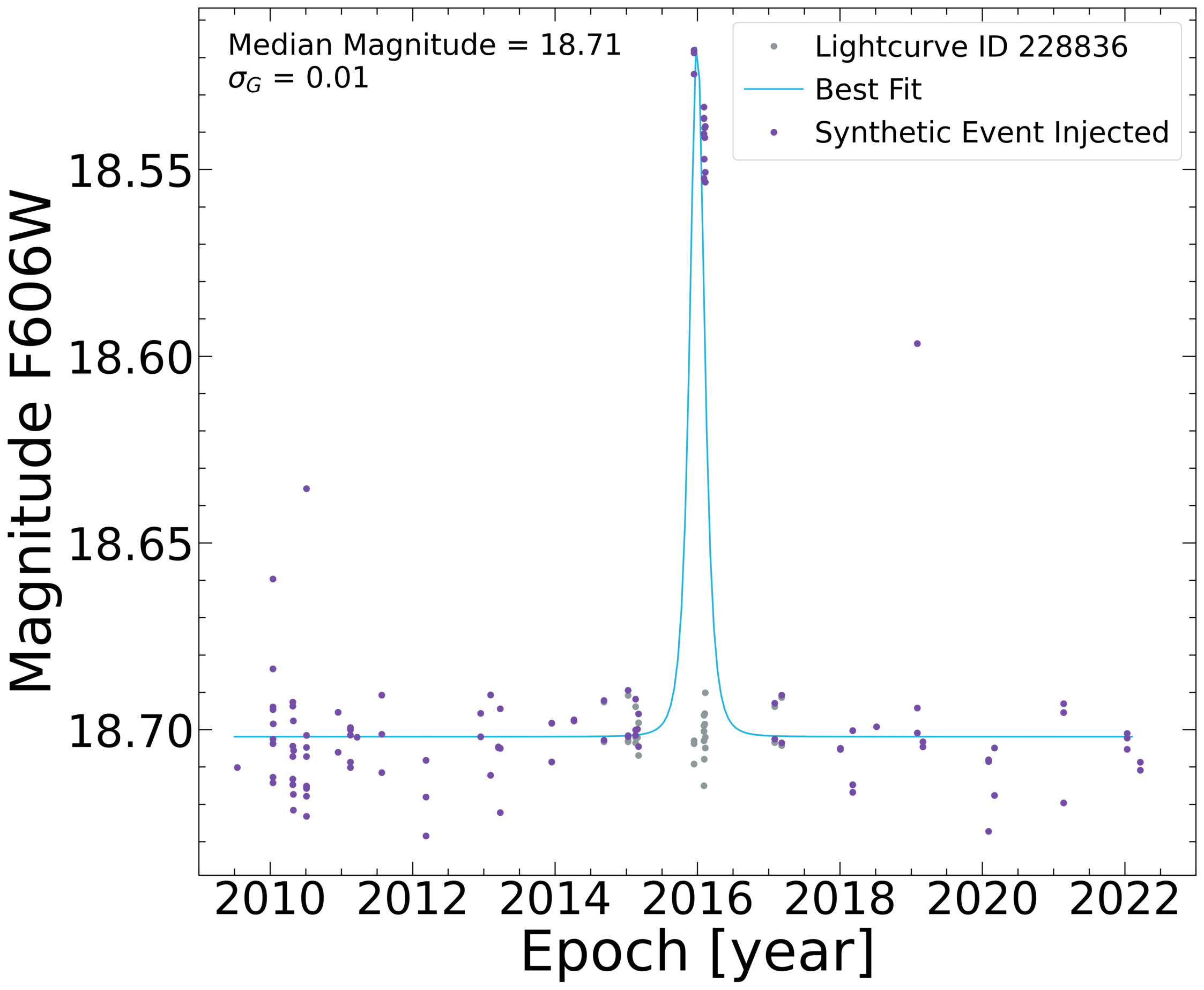}
    \caption{Lightcurve ID 228836 shown as the gray points. This lightcurve has a median magnitude =  18.71 and a $\sigma_G$ = 0.01. The purple points show lightcurve ID 228836 with an injected synthetic lensing event. Injected values include a maximum magnification of 1.2, epoch of maximum magnification of 2016, $v_T$ equal to 16.8 km/s, and $D_{LS}$ equal to 4 parsecs (pc). The blue line shows the best fit to the synthetic lightcurve.}
    \label{fig:id_228836_synth_inj}
\end{figure}

We create realistic synthetic BH microlensing events for two purposes: (1) to help optimize our model fitting procedure to detect microlensing events (Section \ref{subsec:fitting}), and (2) to quantify the sensitivity and completeness of our observations to events with varying parameters (Section \ref{sec:synthetic_analysis}). Using the parameter PDFs fit from the simulation data (Section \ref{subsec:cluster_simulations}), we create a set of 40,000 synthetic microlensing events. These synthetic microlensing events are then added to the data. We inject events with peak magnifications greater than $1.01$ and with epochs of max magnification ranging between 2010.5 and 2021. We characterize the duration of each event by the length of time where the magnification is greater than 1.01; the mean duration is 1.22 years, with a standard deviation of 1.06 years. The distribution of injected maximum magnifications as a function of the duration of the event can be seen in Fig.~\ref{fig:mag_vs_duration}, where we define the fit and injected baseline subtracted maximum magnifications as $\Delta \mu_{fit} = \mu_{fit} - 1$ and $\Delta \mu_{inj} = \mu_{inj} - 1$, respectively. Fig.~\ref{fig:id_228836_synth_inj} shows an example of one synthetic lightcurve. The distributions of the injected parameters can be seen in the bottom panels of Fig.~\ref{fig:parameter_distributions}.

Since we know the values of the injected parameters (i.e., injected maximum magnification, injected epoch of max magnification, etc.), we test our model fits on these synthetic lightcurves and see how well the values are recovered (see Section \ref{sec:synthetic_analysis}).

\subsection{Fitting Lensing Signatures}\label{subsec:fitting}

We use the magnification equation (equation~(\ref{eq:magnification})) to create a model that is then fit to a lightcurve (either observed or synthetic). We rewrite the lens equations (equations~(\ref{eq:y_scaling}) - (\ref{eq:lens_eq})) to depend on three independent parameters: the maximum magnification (corresponding to the minimum y in equation~(\ref{eq:y_scaling})), the epoch of that maximum magnification, and the relative proper motion. This last parameter controls the duration of the event and depends on $v_T$ and $D_L$. In this model, the source and observer are considered to be at rest. To reduce the number of parameters that need to be adjusted when fitting the model to a lightcurve, we set the $D_{LS}$ and $M$ to the median values of the synthetic dataset, \synthdls{} and \synthbhmass{}, respectively. The lensing signature adds to the existing brightness of the star, which we set in our model to be the median magnitude of the lightcurve (see Fig.~\ref{fig:id_228836_synth_inj}).

The fitting process can be divided into two steps: (1) providing initial guesses for the parameters, and (2) obtaining the best-fitting model to the lightcurve through minimizing a cost function. We optimized both of these steps by evaluating synthetic events, maximizing completeness, and optimizing the recovery of input lensing events (Section \ref{sec:synthetic_analysis}).  We detail both of these steps below.  

\subsubsection{Providing Initial Parameters}\label{subsubsec:initial_params}

Before fitting a model to a lightcurve, initial guesses for the parameters must be provided. By providing initial guesses that are close to the best fit, we can both improve the likelihood that a good fit will be obtained and cut down on the number of iterations needed to find the best-fit model, thus reducing computational time. 

For our initial parameter guesses, we focused on robustly estimating the magnification and epoch of maximum magnification of the most likely potential lensing event in any given lightcurve. Due to cosmic rays, diffraction spikes, and uncorrected bad pixels, single bright observations often are spurious signals. To overcome this issue, we performed a four-tier approach, which focuses on identifying groups of bright observations that are expected to be present during a true microlensing event. If an initial parameter guess is not provided in an earlier tier, we fall back to the next tier; earlier tiers are stricter and provide the best initial parameter guesses, while later tiers are more relaxed. We describe each tier below.

\begin{enumerate}
    \item Tier one: we first determine all bright observations with magnitudes more than $2\sigma_G$ brighter than the median magnitude. Then we start with the brightest observation and determine the number of bright observations within $\pm0.1$ years of that observation's epoch. If the number of bright observations within the 0.2-year range is $\geq 3$, we consider it to be a \textit{bright observation cluster} and keep track of the epoch of the brightest member of the cluster. We then progress through each bright observation (excluding observations within the previous bright observation counts) and repeat this step. In the case of a single bright observation cluster, the epoch of its brightest member is used as the initial guess for the epoch of maximum magnification. If multiple clusters are identified, each cluster's brightest member provides a separate initial guess for the epoch of maximum magnification, and the regression is performed independently for each; the best-fit model is then selected by keeping the minimum reduced chi-square value. 
    \item Tier two: if tier one does not produce a bright observation cluster, we then repeat the same procedure with relaxed constraints. Namely, we consider a bright observation to be an observation that is $1.5\sigma_G$ brighter than the median magnitude (rather than $2\sigma_G$), we widen the range from $\pm0.1$ years to $\pm0.15$ years, and also reduce the threshold number of bright observations from 3 to 2. As with tier one, the epoch of the brightest cluster member is used as the initial guess for the epoch of maximum magnification, with multiple clusters handled in the same manner.
    \item Tier three: if tier two is unable to produce an initial parameter guess, we then search for a single bright observation ($2\sigma_G$ brighter than the median magnitude) that does not have dim neighbors ($2\sigma_G$ fainter than the median magnitude) within $\pm0.15$ years of its epoch. A dim neighbor near a bright observation suggests that the bright observation may be due to noise, rather than being a member of a potential microlensing event. If this produces a single candidate bright observation, we use its epoch as the initial guess for the epoch of maximum magnification. If multiple candidate bright observations are produced, we use the brightest observation's epoch as the initial guess for the epoch of maximum magnification.
    \item Tier four: finally, if tier three is unable to determine an initial parameter guess, we choose the brightest observation in the lightcurve and use its epoch as the initial guess for the epoch of maximum magnification.
\end{enumerate}

With the initial epoch guess(es) determined, we then determine the initial guess for the maximum magnification, $\mu$. Magnification values are ratios of fluxes, i.e., $\frac{F_{outlier}}{F_{median}}$. Since we have the corresponding observed magnitudes, we can easily convert to flux and therefore calculate an estimate for the magnification initial guess:

\begin{equation}
    \mu = \frac{F_{outlier}}{F_{median}} = 10^{\frac{m_{median} - m_{outlier}}{2.5}}
\end{equation}

For our initial guess of the relative proper motion, we provide a constant value -- equal to the median of the synthetic dataset's injected relative proper motion -- of \synthpm{}. We derive this value using the median injected transverse velocity (Fig.~\ref{fig:parameter_distributions}) for the 40,000 synthetic lightcurves (Section \ref{subsec:synthetic_events}) and the distance to \ocen{} (5,430 pc, as seen in \citealt{Baumgardt_2021}).

\subsubsection{Fitting a Lensing Model: a Physically Motivated Cost Function}\label{subsubsec:fitting_model}

To detect potential lensing events, we optimize a fitting method that finds the best-fit lens model for each source, with the goal of recovering the epoch, maximum magnification, and duration of the event. 

We know from our synthetic lightcurve simulation data that high magnification events are rare relative to lower magnification events (Fig.~\ref{fig:parameter_distributions}). Using this information, we can incorporate this into our fitting method and use a Maximum a Posteriori (MAP) estimate to obtain our best-fit model. We add a prior term to our cost function and rewrite this as the log-posterior distribution function of the parameters ($h$) given the data ($D$):

\begin{equation} \label{eq:posterior}
    P(h | D) = p(D | h) + p(h)
\end{equation}

The two terms are a likelihood function:

\begin{equation} \label{eq:likelihood}
    p(D | h) = \frac{1}{2\sigma_G^2} \Sigma_{i = 0}^{n} [\sqrt{w_i}(y_{i} - y^{\widehat{}}_{i})]^{2}
\end{equation}

and a prior based on the log-normal distribution of synthetic events:

\begin{equation} \label{eq:prior}
    p(h) = \ln(\mu_{fit}) + \frac{(\ln (\mu_{fit}) - \mu_M)^2}{2\sigma_M^2 }
\end{equation}

In equation~(\ref{eq:prior}), $\mu_{fit}$ is the magnification that is fit, while $\mu_M$ and $\sigma_M$ are the mean and standard deviation of the underlying normal distribution of the log-normal prior distribution, respectively. The values of $\mu_M$ and $\sigma_M$ are derived from the log-normal fit reported in Table~\ref{table:dist_params}; specifically, $\mu_M$ is the natural log of the scale parameter and $\sigma_M$ is the shape parameter of the fitted log-normal distribution (where the scale and shape parameters are SciPy specific \citep{scipy_2020}), and are therefore distinct from the mean and standard deviation of the log-normal distribution reported in the table. Note that constant terms have been dropped since they do not affect optimization.

Both equation~(\ref{eq:likelihood}) and equation~(\ref{eq:prior}) have been multiplied by $-1$, allowing us to minimize equation~(\ref{eq:posterior}). Note that we only use an informative prior for the magnification, while for other parameters we use uniform priors and therefore they do not appear in equation~(\ref{eq:prior}).

In equation~(\ref{eq:likelihood}), we scale the residuals of observations flagged by the extended diffraction spike mask (Section \ref{subsubsec:diff_spikes}) by a factor of $\frac{1}{8}$ (i.e., when the observation $y_i$ is flagged, $w_i = \frac{1}{8}$, otherwise $w_i = 1$). By performing this scaling, the flagged observations contribute less to the cost function, and therefore are not prioritized during fitting, but still provide some signal.

\section{Synthetic Lightcurve Analysis}\label{sec:synthetic_analysis}

Using our synthetic lightcurves (Section \ref{subsec:synthetic_events}), we quantify our detection completeness to better understand which events can be recovered from observed data and to optimize our model for maximum recovery. Our synthetic lightcurves also enable the development of binary classification models that label a lightcurve as positive if it contains a microlensing event and negative otherwise. Using the post-regression statistics from the MAP model fits, we either apply selection thresholds directly (e.g., a lightcurve is labeled positive if a statistic is greater than some threshold(s)) or use the statistics as input features for a neural network. Completeness analysis and binary classification training using synthetic lightcurves are detailed in Sections \ref{subsec:completeness} and \ref{subsec:binary_classification}, respectively.

\subsection{Completeness}\label{subsec:completeness}

To determine our ability to detect potential microlensing events, we assess the completeness of our dataset using synthetic events.  

We consider a synthetic event detected based on two possible conditions: (1) testing if the best fit model's fit epoch of maximum magnification is within the duration of the injected synthetic microlensing event, and (2) requiring an additional constraint that the best fit model's $\Delta \mu_{fit}$ is within 80\% of $\Delta \mu_{inj}$. Note that we define the duration of the injected synthetic microlensing event as the time span where the magnification is greater than 1.01. 

\begin{figure} 
    \centering
    \includegraphics[width=1\linewidth]{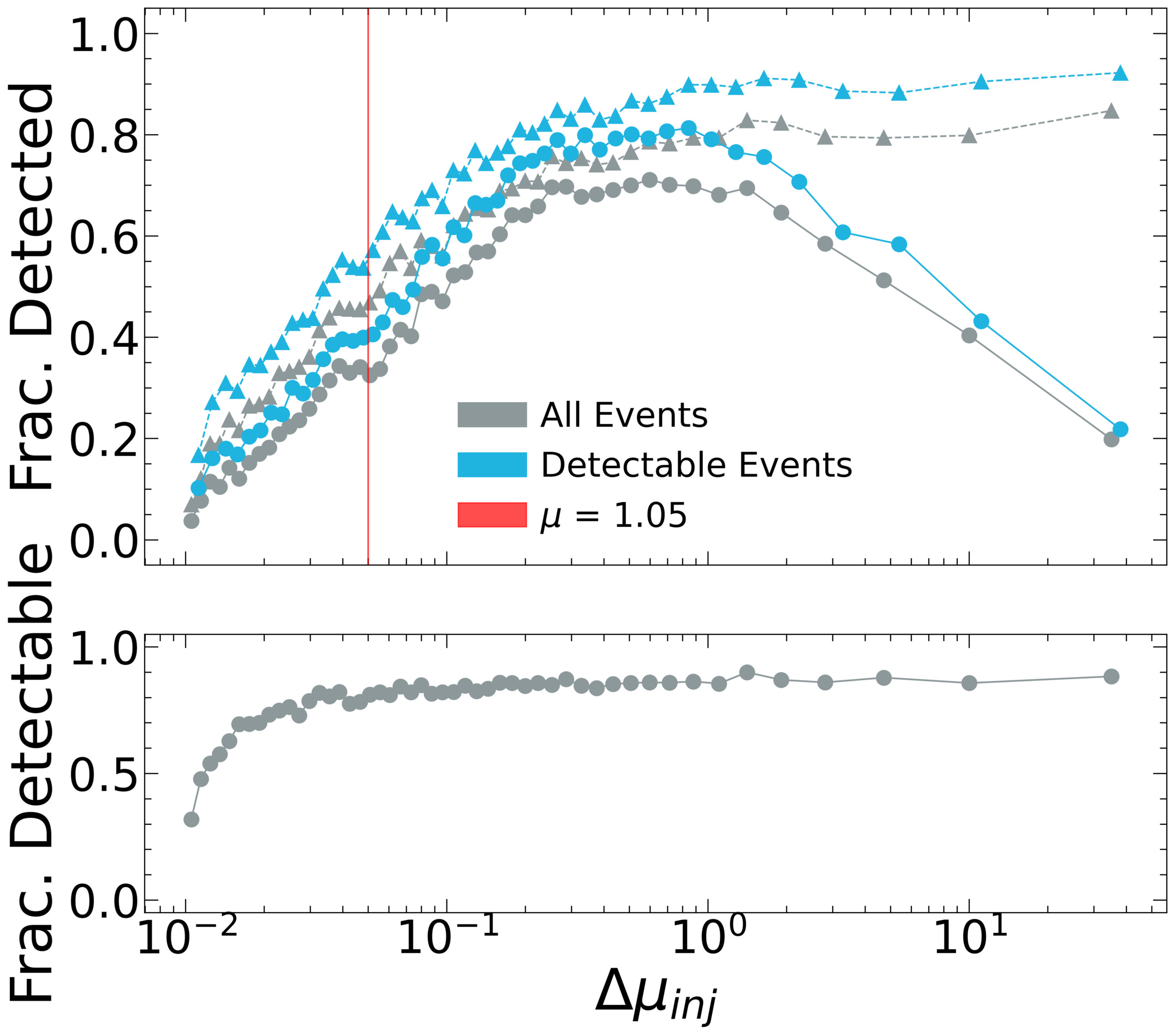}
    \caption{In the top panel, we show the fraction of recovered events as a function of the injected baseline subtracted maximum magnification. The dashed lines with triangle markers represent events where the epoch of the event was correctly recovered, while the solid lines with circle markers represent where both the epoch was recovered and where the fit baseline subtracted maximum magnification was within 80\% of the injected baseline subtracted maximum magnification. The gray lines calculate the fraction of recovered events with respect to the total number of synthetic lightcurves (40,000) while the blue lines calculate the fraction of recovered events with respect to the subset of synthetic lightcurves considered to be detectable (31,640). Each line has 50 points; each point in the gray lines contains 800 synthetic lightcurves, while each point in the blue lines contains roughly 630 synthetic lightcurves. The injected baseline subtracted maximum magnification is the median value of the bin members. In the bottom panel, we show the fraction of events considered detectable as a function of the injected baseline subtracted maximum magnification. Each point contains 800 synthetic lightcurves.}
    \label{fig:magnification_completeness}
\end{figure}

Fig.~\ref{fig:magnification_completeness} shows the fraction of sources detected as a function of the injected baseline subtracted maximum magnification. In some cases, the duration of the injected synthetic microlensing event falls in gaps in our observations. Therefore, for assessing completeness, we also show completeness just for synthetic lightcurves with at least two observations during the microlensing event duration; these are labeled ``Detectable Events'' and shown in blue.
The fraction of detected synthetic events increases by nearly 10\% across the board when considering detectable synthetic events. In the bottom panel, the fraction of sources that are considered detectable is shown as a function of the injected baseline subtracted maximum magnification; a total of 31,640 synthetic lightcurves contain detectable synthetic events out of a total of 40,000 synthetic lightcurves.

\begin{figure}
    \centering
    \includegraphics[width=1\linewidth]{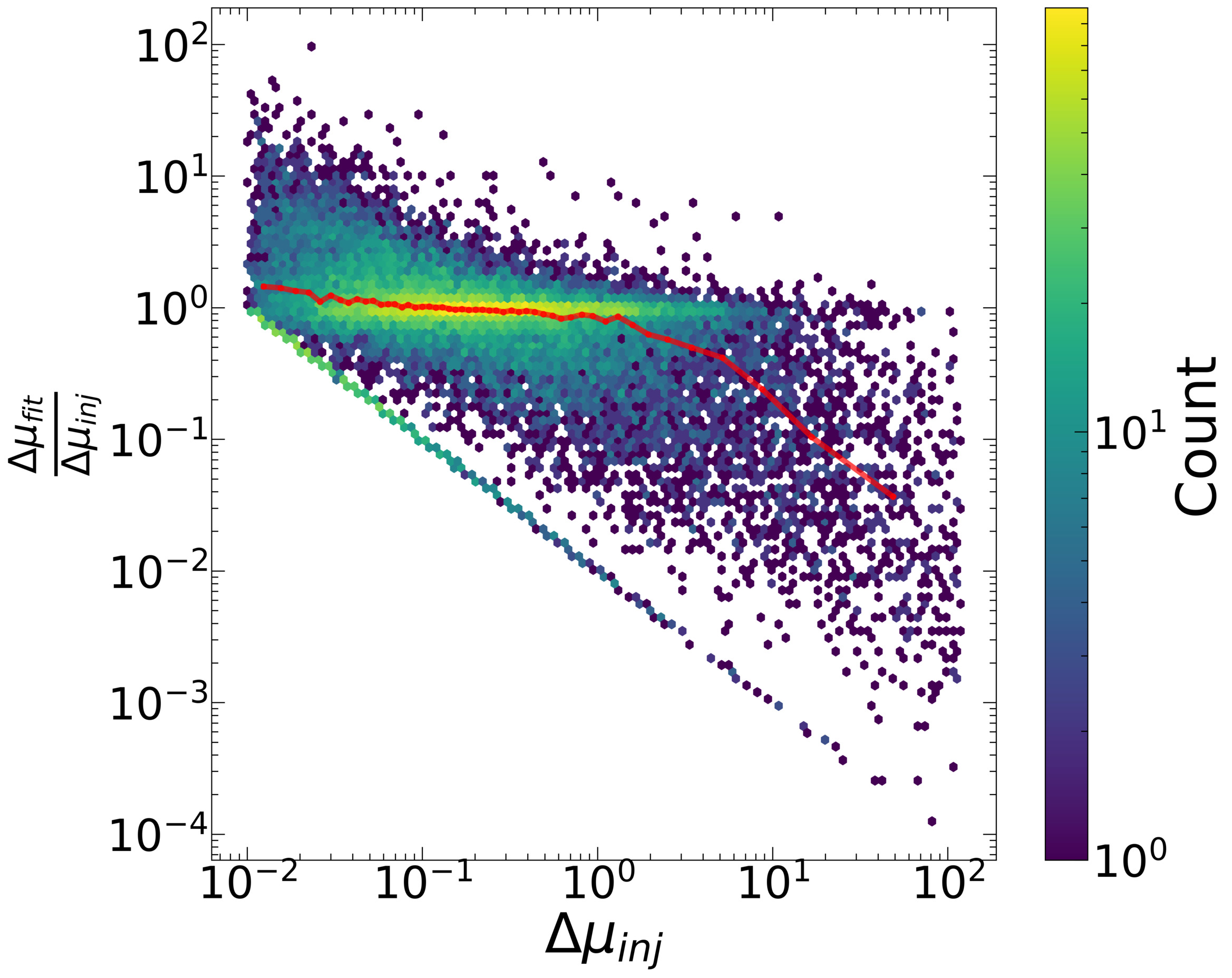}
    \caption{The ratio of baseline subtracted maximum magnifications (fit / injected) versus the injected baseline subtracted maximum magnification of synthetic lightcurves where the epoch of the event was correctly recovered (22,156 / 40,000 synthetic lightcurves). At small injected magnifications, the ratio tends to be greater than 1, while at larger magnifications it drops below 1. The red line represents the median values; each marker on the line contains roughly 440 lightcurves in the bin. The cut off towards the bottom of the data shown is where the fit magnification reaches a minimum ($\mu_{fit} = 1.01$).}
    \label{fig:magn_ratio}
\end{figure}

Fig.~\ref{fig:magnification_completeness} shows we recover the epoch of a majority of the events above an injected magnification of $\sim$1.05 ($\Delta \mu_{inj} \sim 0.05$). At large magnifications, the fraction of detected synthetic events begins to drop if we add the constraint that the magnification is recovered ($\Delta \mu_{fit}$ within 80\% of $\Delta \mu_{inj}$); this is due to the limited temporal coverage of the injected microlensing event; frequently only a portion of the maximum magnification of the event is captured and thus the magnification can be significantly underestimated. This limitation is clearly illustrated in Fig.~\ref{fig:magn_ratio}, where, as one goes to larger $\Delta \mu_{inj}$, the ratio $\frac{\Delta \mu_{fit}}{\Delta \mu_{inj}}$ tends to decrease, as shown by the median line. This is related to the penalty placed on large magnifications in our cost function; however, this penalty reduces the over-estimate of the magnification at lower magnifications where we expect most of our true lensing events.

\subsection{Binary Classification Modeling}\label{subsec:binary_classification}

To gain a better understanding of what events we can recover and to determine whether an observed lightcurve truly contains a microlensing event, we use two approaches for converting our regression problem into a binary classification problem: (1) applying thresholds directly to interpretable statistics -- for example, the reduced chi square of the model fit -- to assign binary labels, and (2) training a neural network to predict the probability of an event being within a lightcurve, where we can then apply a threshold to recover binary labels. We refer to instances that are predicted to contain a lensing event as a \textit{positive event}; otherwise, we refer to it as a \textit{negative event}.

To train our models, we create 172,046 synthetic lightcurves -- the same number of observed lightcurves that we have in our final dataset (Section \ref{subsubsec:final_dataset}) -- without a synthetic microlensing event (our \textit{negative events}) in addition to the 40,000 synthetic lightcurves with synthetic microlensing events (our \textit{positive events}) described in Section \ref{subsec:synthetic_events} and fit a lensing model (Section~\ref{subsubsec:fitting_model}) to each lightcurve. 

We divide the data into training, validation, and testing sets using a stratified random split to maintain class balance, with 80\% of the data used for training, 10\% of the data used for validation, and the remaining 10\% used for testing. This training split allows most of the data to be used for training, while having reasonable amounts of data for validation and testing. The validation set is used to calibrate thresholds for both approaches and compare the performance of models, while the test set is used purely to evaluate the model that performed best on the validation set. This ensures that data leakage is minimized, models are fairly evaluated against each other, and that they perform as expected on unseen data.

Since we expect only 1-2 true microlensing events in the data, we want as few false positives (synthetic lightcurves without events predicted by the model to have events) as possible while retaining as many true positives (synthetic lightcurves with events predicted by the model to have events) as possible. We therefore score models by the \textit{True Positive Rate (TPR)} at a \textit{False Positive Rate (FPR)} of 0.01\% (i.e., the model with the largest TPR at an FPR $\leq0.01\%$ is the best model). The outline of training, validation, and testing can be seen below. 

\subsubsection{Threshold-Based Classification}\label{subsubsec:thresholding}

We started by applying a threshold to the difference in reduced chi square values between the constant line fit and the model fit, $\chi^2_{c-m}$, where $\chi^2_{c-m} = \chi^2_{red., constant} - \chi^2_{red., model}$ and the subscript \textit{red.} denotes that it is the reduced chi-square statistic. If a lightcurve contains a microlensing event and is successfully recovered by our model, then this difference should be positive; therefore, any lightcurve with a $\chi^2_{c-m}$ value greater than the tuned threshold is predicted to be a positive event. Since this approach does not involve training a model in the traditional sense of machine learning (i.e., training model weights), we are able to skip the training set and directly evaluate different thresholds on the validation set. We focus on achieving an FPR of 0.01\% as this would result in $\sim$17 false positives in our full dataset.  Using this simple threshold, we achieve a TPR of 5.6\% with a threshold value of 17.57; this can be seen in Fig.~\ref{fig:roc_curve_nn} (gray line), which shows TPR vs FPR, also known as a Receiver Operating Characteristic (ROC) curve.

Using $\chi^2_{c-m}$ alone provides interpretability, but does not use much information. In an attempt to increase the TPR at an FPR of $0.01\%$ while still maintaining the interpretability of this approach, we introduced an additional threshold using the number of bright observation clusters (as defined in Section \ref{subsubsec:initial_params}), where a lightcurve is predicted as a positive event when the number of bright clusters is greater than 0 and $\chi^2_{c-m} > 14.1$ (as before, thresholds tuned at an FPR of $0.01\%$). However, this yielded negligible improvement over the single-statistic thresholding at the FPR of interest, and is omitted from Fig.~\ref{fig:roc_curve_nn} for clarity.

\subsubsection{Multi-Statistic Thresholds} 

In this subsection, rather than using a single threshold, we look at threshold based candidates using multiple statistics. We focus on four statistics expected to carry high discriminating power: the reduced chi square of the model fit $\chi^2_{red., model}$; the reduced chi square difference $\chi^2_{c-m}$, which captures the difference between the constant line fit and the model fit reduced chi square values; the number of dim outliers within the lensing fit plus the number of bright outliers outside the lensing fit (observations $2\sigma_G$ fainter or brighter than the lightcurve's median magnitude, respectively), $n_{dim-fit} + n_{bright-out}$; and $n_{bright-fit}$, the number of bright outliers within the lensing fit. 

We apply thresholds directly to the statistics themselves by computing a 2D histogram from two statistics derived from synthetic and observed lightcurves, counting the number of each in every bin ($n_{synth}$ and $n_{obs}$). For each bin, we calculate the ratio $n_{synth} / (n_{synth} + n_{obs})$, which serves as an ad-hoc likelihood that an observed lightcurve in that bin represents a lensing event. A ratio near 1 suggests the observed lightcurve closely resembles synthetic ones, implying a higher likelihood of lensing, while a ratio near 0 indicates that the bin is dominated by observed lightcurves, implying a lower likelihood of lensing. We then apply a single threshold analogous to the direct cuts in Section \ref{subsubsec:thresholding}. In both of the techniques described below, we apply a Gaussian kernel with a standard deviation of 1 to smooth the features.

\begin{figure}
    \centering
    \includegraphics[width=1\linewidth]{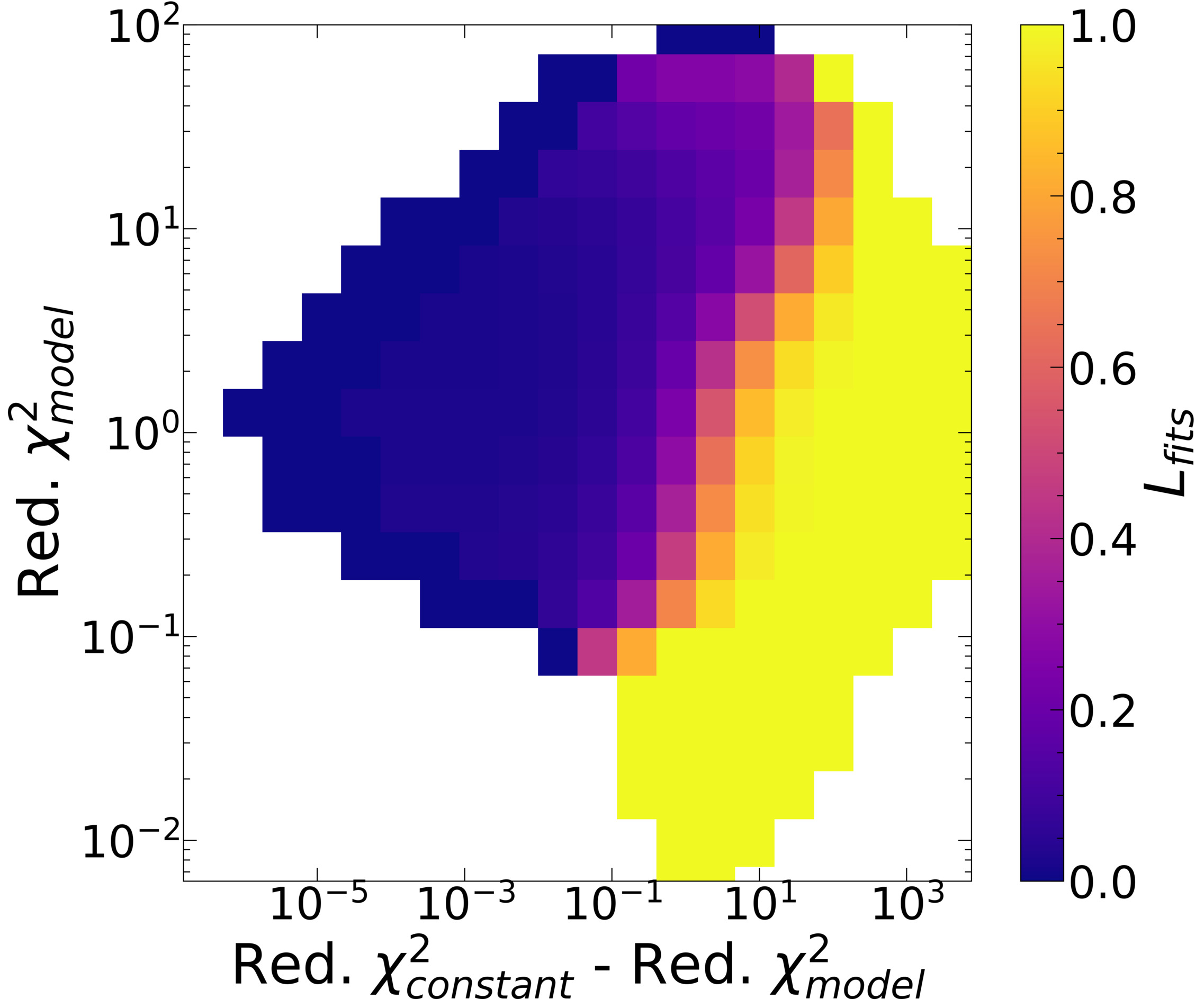}
    \caption{The difference in reduced chi squares between a constant line fit and the model line fit vs. the reduced chi square of the model line fit. The colorbar shows $L_{fits}$.}
    \label{fig:chi_diff}
\end{figure}

We first apply this technique using the statistics $\chi^2_{c-m}$ and $\chi^2_{red., model}$. When $\chi^2_{c-m}$ is positive, the model has a statistically better fit than the constant line. In Fig.~\ref{fig:chi_diff}, the highest likelihood values are where this difference is positive, as expected. We refer to this ad-hoc likelihood as $L_{fits}$.

\begin{figure}
    \centering
    \includegraphics[width=1\linewidth]{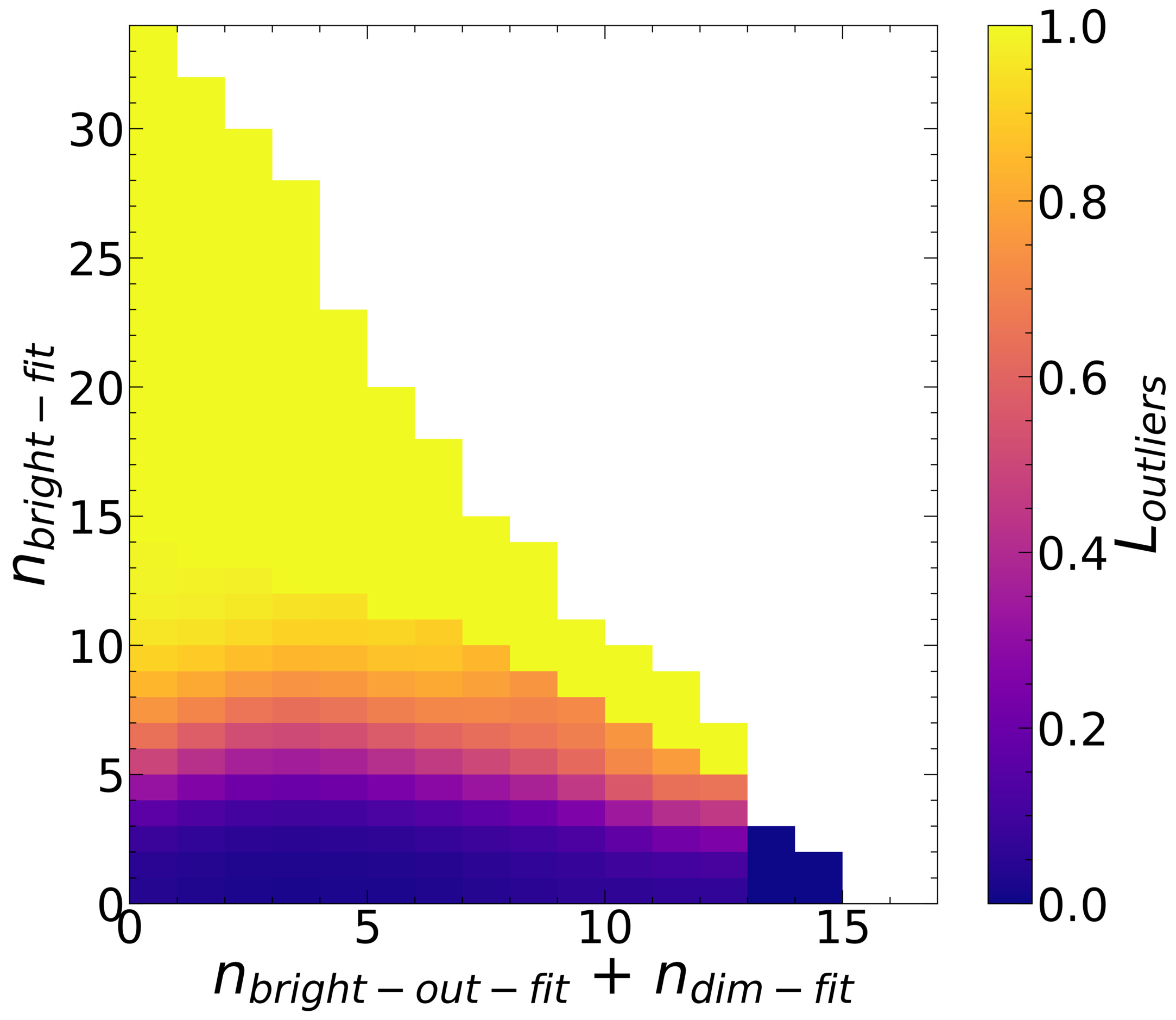}
    \caption{The number of outliers ($|N_{i,j}| \geq 2$) in three regions of the lightcurves. The colorbar shows $L_{outliers}$.}
    \label{fig:outlier_ratios}
\end{figure}

We also apply this method to $n_{dim-fit} + n_{bright-out}$ and $n_{bright-fit}$. Lightcurves containing potential microlensing events will ideally have small values for the former, while the latter should have large values. As seen in Fig.~\ref{fig:outlier_ratios}, observed lightcurves in the left portions of the plot have the highest likelihoods, as expected; regions colored bright yellow are regions where there are only synthetic lightcurves in the bins and no observed lightcurves. We refer to this ad-hoc likelihood as $L_{outliers}$. 

Using these two ad-hoc likelihoods, we then multiply them together to help identify potential events in observed lightcurves (i.e., $L_{fits} * L_{outliers}$). However, when inspecting the most likely ($L_{fits} * L_{outliers} > 0.3$; $n=62$) candidates, we found that nearly every candidate was a blended source; i.e., there was a neighboring source within a couple pixels. Blending mimics the statistical signature of lensing events, which explains why these statistics in isolation are insufficient for reliable event recovery. This demonstrates that while $\chi^2_{c-m}$, $\chi^2_{red., model}$, $n_{dim-fit} + n_{bright-out}$, and $n_{bright-fit}$ carry discriminating power, the neural network's ability to leverage the joint distribution of many statistics is what enables it to distinguish genuine lensing candidates from contaminants such as blended sources.

\subsubsection{Neural Network Classification}\label{subsubsec:nn_classification}

Although applying thresholds directly to post-regression statistics is interpretable, it uses limited information for classification. We could increase the number of statistics to which we apply thresholds, but this begins to become less interpretable and is a complicated space to explore. To solve these problems, we train a neural network on many statistics that predicts the probability that a microlensing event is in a lightcurve. We can then apply a single threshold to the output probabilities from the neural network while using many statistics as input features to the neural network (see Table~\ref{table:nn_features} for details on the inputs to the neural network). 

\textbf{Architecture and Training.} We train a neural network following standard training protocols, i.e., train with the training set, use the validation set to track generalization, tune hyperparameters, and calibrate the threshold, then assess the neural network's performance with the best set of hyperparameters on the test set; see Appendix Section \ref{app:nn_training} for more details. 

\begin{figure}
    \centering
    \includegraphics[width=1\linewidth]{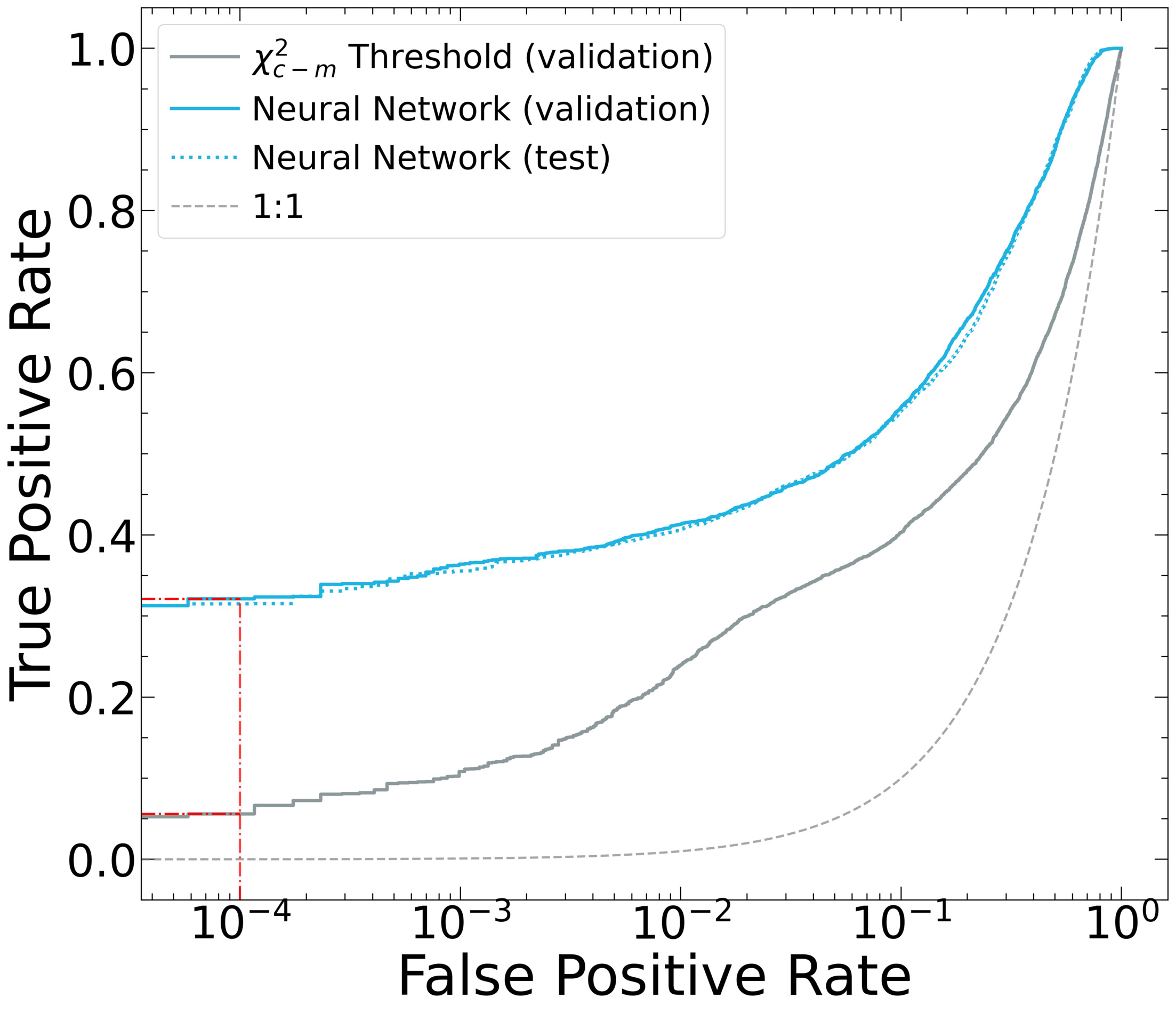}
    \caption{A comparison of ROC curves between a neural network classifier, the blue lines, and the statistic threshold based classification (Section \ref{subsubsec:thresholding}), the gray line. We compare approaches using the validation set (solid lines) and show our best model (neural network) test set performance (dashed blue line) for completeness. The neural network's test set ROC curve is nearly identical to the validation set ROC curve. The vertical red dash-dot line marks a $0.01\%$ FPR, while the horizontal red dash-dot lines mark the statistic threshold based classification TPR ($5.6\%$) and neural network TPR ($32.13\%$).}
    \label{fig:roc_curve_nn}
\end{figure}

\textbf{Validation and Recovery Rate.} At an FPR of $0.01\%$, the best neural network achieves a TPR of $32.13\%$ with a probability threshold of $98.69\%$ on the validation set (i.e., if the neural network outputs a probability of $98.69\%$ or greater for a given lightcurve, it is predicted to contain a lensing event). The neural network performs significantly better than the single threshold on the $\chi^2$ difference described above, providing an increase of approximately $27\%$ on the TPR at an FPR of $0.01\%$, as seen in Fig.~\ref{fig:roc_curve_nn}. This justifies the loss of interpretability of our neural network model.
 
We are interested in understanding how the threshold determined on the validation set performs on unseen data, i.e., the test set, as a proxy for how the network will perform on observed data. The FPR remains unchanged at $0.01\%$ and the TPR decreases by less than a percent to $31.5\%$; because this difference is negligible, we do not mark the TPR of the test set separately in Fig.~\ref{fig:roc_curve_nn}. This indicates that the validation-tuned threshold generalizes well to unseen synthetic data. It is important to note, however, that the test set is drawn from the same distributions as the training and validation sets, so similar performance is expected. The more significant concern is the shift in data distribution when applying the network to observed lightcurves, which we discuss further in Section \ref{sec:bh_pop_disc}. 

\begin{table}
	\centering
	\caption{The performance of the neural network classifier on the validation set and test set. We determined the threshold on the validation set that provided an FPR of 0.01\% and evaluated on the test set to see how the FPR and TPR would change on unseen data. As seen, the change is negligible, with less than a percent drop in the TPR.}
	\label{table:nn_metrics}
	\begin{tabular}{cccc}
		\hline
		Dataset & FPR & TPR & Threshold\\
		\hline
		Validation & 0.01\% & 32.13\%  & 98.69\%\\
        Test       & 0.01\% & 31.5\%   & 98.69\%\\
		\hline
	\end{tabular}
\end{table}

\begin{figure}
    \centering
    \includegraphics[width=1\linewidth]{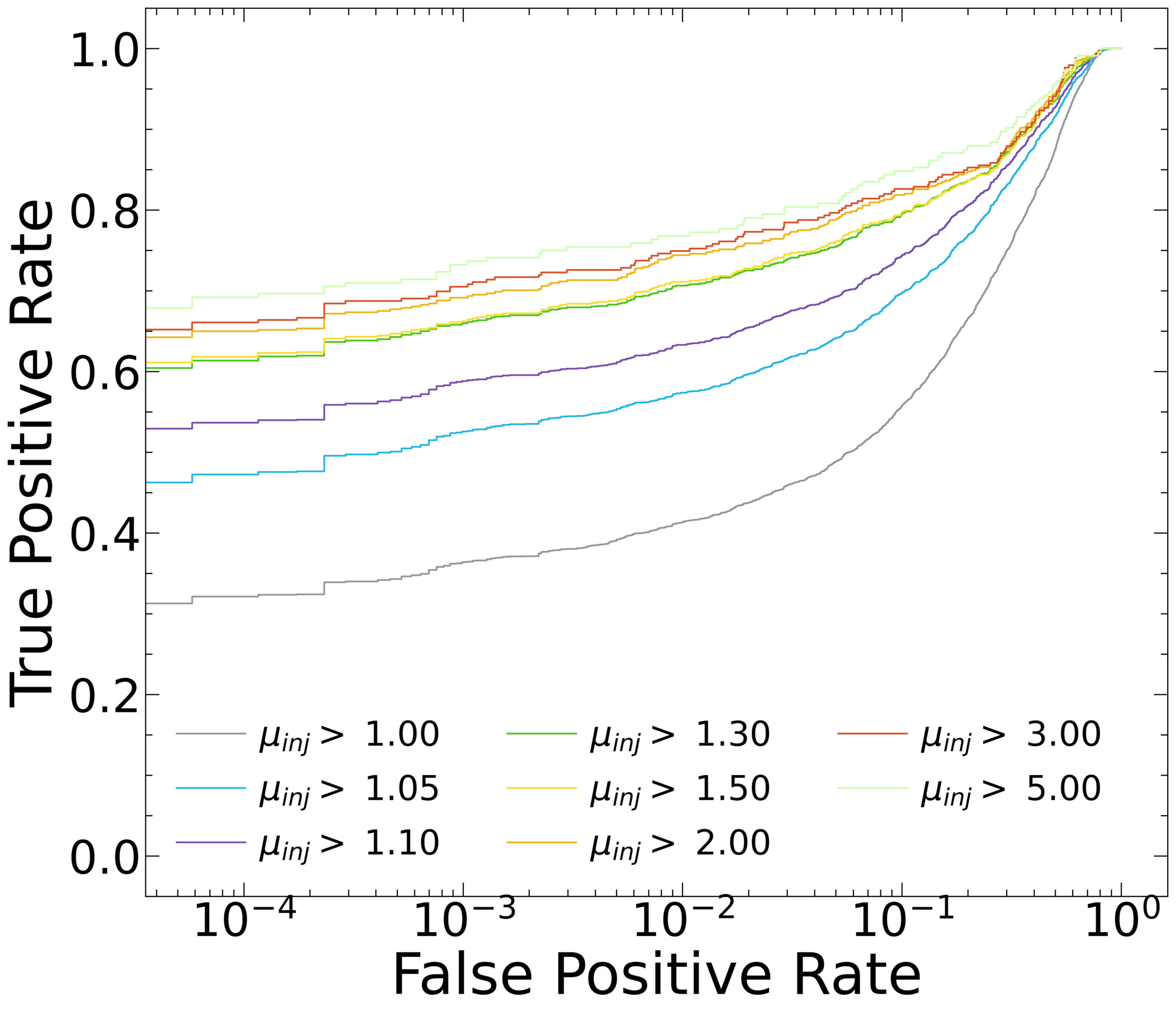}
    \caption{ROC curves of subsets of the validation set with varying magnification values. Note that the magnification threshold applies only to the positive set of synthetic lightcurves (i.e., those with an injected lensing event); the negative lightcurves remain unchanged throughout.}
    \label{fig:roc_mag_subsets}
\end{figure}

\textbf{Recovery Rate as a Function of Magnification.} To understand if there are magnifications of microlensing events that we are certain to recover with our neural network, we inspect ROC curves on the validation set using a subset of positive examples with injected synthetic microlensing events above certain magnifications. The results can be seen in Fig.~\ref{fig:roc_mag_subsets}; as expected, as the injected magnification threshold increases, the TPR at a given FPR also increases. At an FPR of $0.01\%$, the TPR reaches $70\%$ for our largest magnification subset ($\mu_{inj} > 5$); this suggests that if a high-magnification event is present in the observed lightcurves, we are significantly more likely to recover it.

\section{Candidate Black Hole Lensing Events}\label{sec:bh_pop_disc}

We use the trained neural network (Section \ref{subsubsec:nn_classification}) to assist in identifying potential candidates. Using the validation tuned threshold of $98.69\%$, we find 1,213 potential lensing candidates in the observed lightcurves. This number of candidates is much larger than we would expect, given that the threshold was derived at an FPR of $0.01\%$, which would result in $\sim17$ candidate lensing events. As discussed in Section \ref{subsubsec:nn_classification}, moving from synthetic data -- which the neural network was trained on -- to observed data showcases a significant shift in the data distribution. 
We suspect that the data shift is the result of shuffling between epochs in the process of making the synthetics, and that the data has more correlated noise on short timescales.  This is likely the result of two factors: (1) data taken at similar times is often at the same roll angle, and PSF related features from nearby bright stars apart from the diffraction spikes may create correlated signals, and (2) data taken in the same visit may be in similar regions of the detector. 

Due to the large number of candidates, we apply additional criteria to reduce the number of potential candidates (Section \ref{subsec:candidate_selection}) to allow us to rank each individual candidate using a rubric (Section \ref{subsec:rubric_ranking}). In Section \ref{subsec:rubric_ranking}, we highlight our final 51 candidate lensing events; we find no single candidate that has overwhelming evidence of being a lensing event.  In Section \ref{subsec:total_bh_population} we nonetheless place some constraints on the total number of lensing events.

\begin{figure*}
    \centering
    \includegraphics[width=1\linewidth]{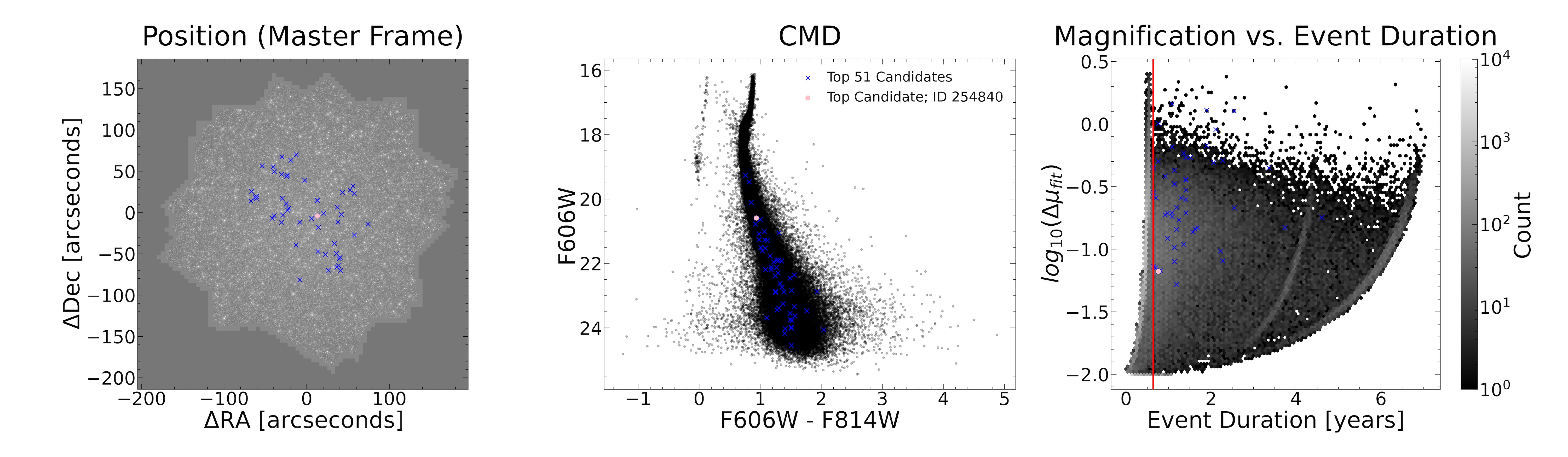}
    \caption{Top Candidate Profile.  Each panel shows the full dataset in grayscale, with the top 51 candidates shown as blue x markers. The top candidate, ID 254840, is shown as a pink circle marker. {\em Left -- } The stacked image of all sources is shown.  {\em Middle -- } A color-magnitude diagram, with a calculated color of F606W-F814W.  {\em Right --}the fit magnification in log-scale as a function of the event duration in years, with the red vertical line marking the 95th percentile of white dwarf lensing durations.  Grayscale here indicates the best-fit values for the entire sample of stars. }
    \label{fig:top_cands}
\end{figure*}

\subsection{Candidate Selection}\label{subsec:candidate_selection}

To reduce the number of candidates, we discard shorter duration events with a fit lensing duration less than 0.64 years, as these may be due to white dwarf lensing events. The ratio of white dwarf and black hole lensing events was previously predicted for two lower mass clusters by \citet{Kiroglus2022}; scaling their results using the \citet{prieto_2025} simulations within the central 90'' of $\omega$~Cen, we expect $\sim$20$\times$ the number of WD microlensing events relative to BH events. Using the white dwarf mass distribution from the \cite{prieto_2025} simulations, we are able to create synthetic microlensing events in the same manner as described in Section \ref{subsec:synthetic_events}. We created 40,000 synthetic lightcurves using the white dwarf mass distribution, and found that these microlensing events are typically shorter than those of BHs. The threshold of 0.64 years eliminates 95\% of all white dwarf events while retaining $\sim$67\% of BH lensing events. In addition to the event duration threshold, we remove all candidates where the fraction of observations flagged by the extended diffraction spike mask (Section \ref{subsubsec:diff_spikes}) within the fit duration is greater than 0.33. In total we are left with 444 candidates above our neural network threshold.  

We then apply these additional cuts to select the highest quality candidates:
\begin{itemize}
\item The $\Delta \mu_{fit}$ should be $>$2$\sigma_G$ of the lightcurve median.
\item That at least one positive outlier with at least 2$\sigma_G$ is present within the fit window.
\item That the contribution of neighboring stars within a 0$\farcs$05 radius aperture is $<$50\% of the flux.
\item That both the {\sc qfit}, which measures the quality of the PSF fit, and {\sc RADXS}, which measures the excess of flux with respect to the PSF prediction, are not more than 4$\sigma$ outliers relative to other stars with similar magnitude.  
\end{itemize}

This leaves us with a list of 51 candidates. These are highlighted in Fig.~\ref{fig:top_cands}. These candidates favor the inner $\sim$90 arcseconds over which the time coverage is best.  They are all main sequence stars, roughly following the distribution of stars in the cluster. In the right panel of Fig.~\ref{fig:top_cands} we can see the top candidates sample from the distribution of best-fit lensing models magnification and event durations.  This distribution also mimics the distribution of synthetic BH events shown earlier in Fig.~\ref{fig:mag_vs_duration}.  

\subsection{Rubric and Visual Ranking of Candidates}\label{subsec:rubric_ranking}
\begin{figure*}
    \centering
    \includegraphics[width=0.48\linewidth]{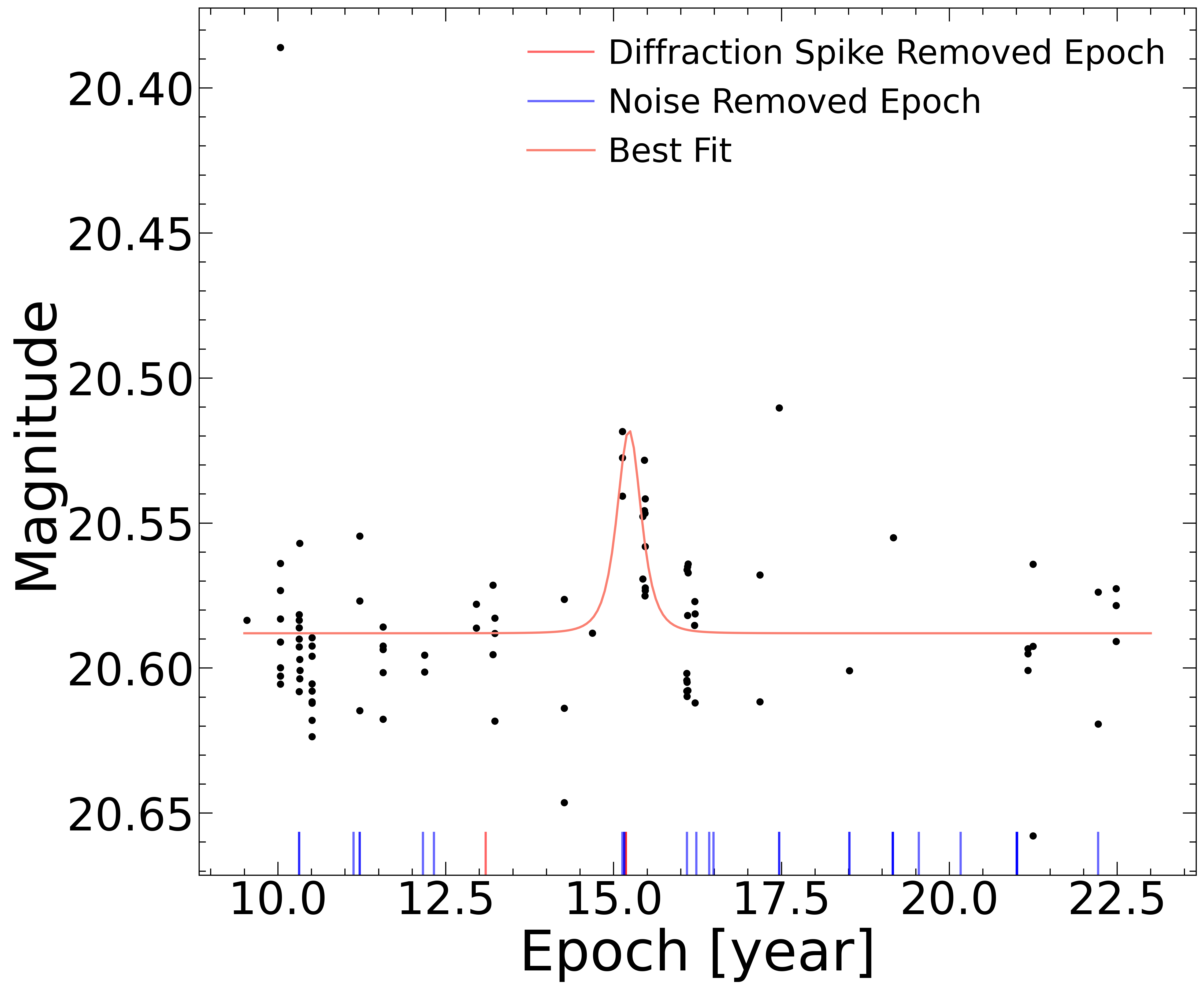}
    \raisebox{1cm}{\includegraphics[width=0.335\linewidth]{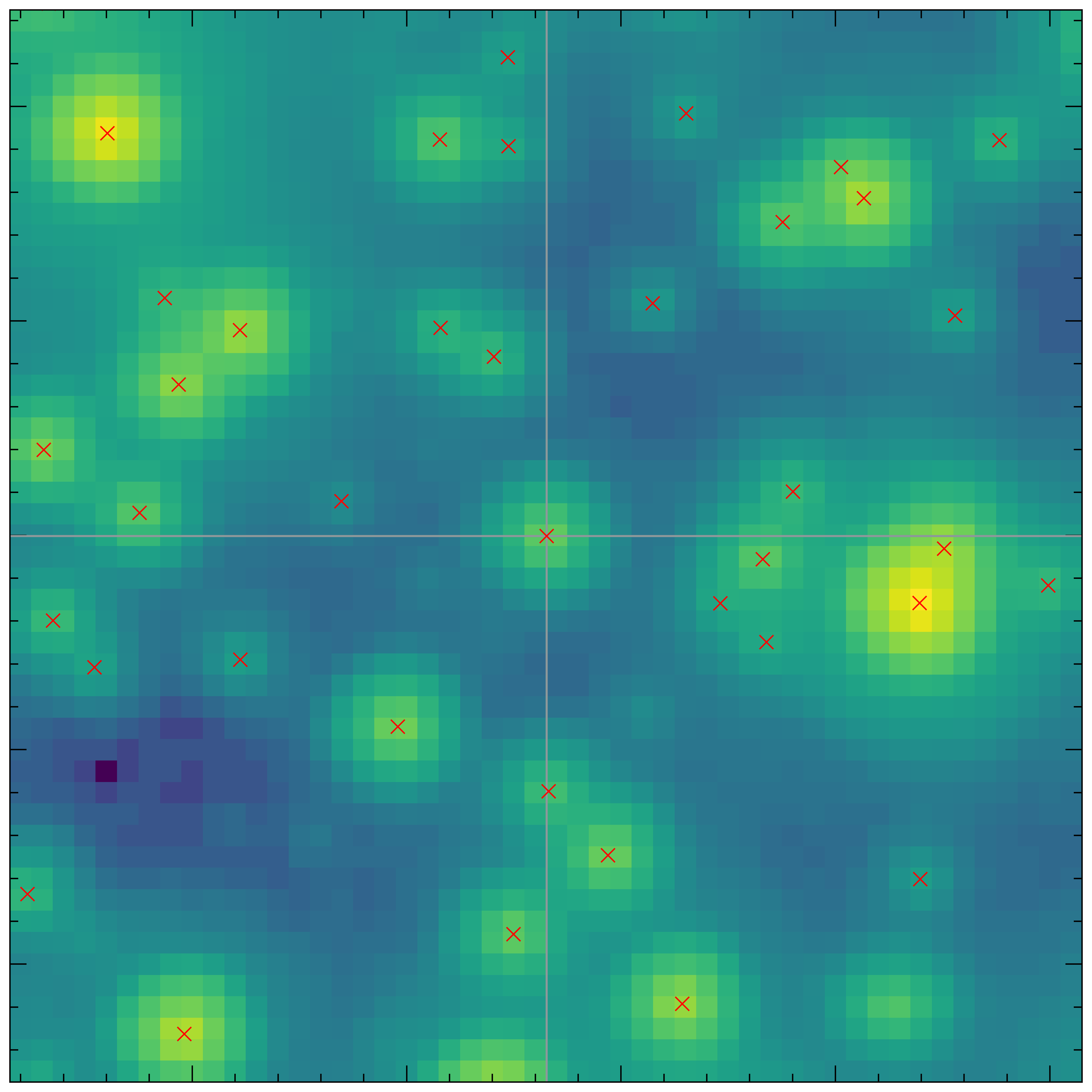}}

\caption{Candidate Microlensing Event.  {\em Left --} Lightcurve of Star 254840 (black points, photometry method 1) including the best fit lightcurve (red). Flagged data are shown as red and blue lines on the bottom. {\em Right -- } Image of the candidate showing it is relatively free of nearby contaminants.  All cataloged sources are shown with red Xs, and the background image is a deep stacked image of all F606W exposures.}
\label{fig:best_candidate}
\end{figure*}

With these 51 candidates, three co-authors scored the top 51 candidates to looks for signs that the signal might be some other contaminant other than a true microlensing signal. This score was done using a rubric that included assessment of the {\sc qfit} and {\sc RADXS} photometric parameters and astrometry in each epoch, an assessment of whether the event was present in all three photometry methods, whether the event peak was in a gap in observations, and a visual assessment of the prominence of the lightcurves (including comparison to a synthetic ``twin" lightcurve with the same event inserted), and the possibility that neighboring star contamination might contribute to the brightening of the star mimicking a lensing signature. Each candidate started with a score of 10 and any criteria that the star met (e.g.~it had an astrometric offset during the brighter epochs) counted negatively towards its score. The top ranked candidate ended with an average score of 7.8, the lightcurve and environment of this candidate is shown in Fig.~\ref{fig:best_candidate}. Unlike many of our candidates, this candidate is a fairly bright star and thus has minimal contamination from overlapping nearby stars, and also has no very bright star in the immediate vicinity.  It clearly brightens in all three photometric methods. Its lightcurve also has good coverage during the peak of the data. Many of our candidates had peaks between epochs, had clear evidence of potential contamination from overlapping stars or being in the wings of the PSF of a very bright star.  Despite these positives, the candidate (like all of our candidates) are not convincing microlensing candidates; it had only 5 outliers above 2$\sigma_G$. Furthermore, we found a bias towards the best candidates being found around the 2016 epoch; while there is a lot of data in this area, we did not find a similar bias among synthetic detections.  

To conclude, we do not feel that any single candidate is a compelling BH microlensing detection. Many synthetic observations showed much clearer detections suggesting minimal evidence for any true BH microlensing events. We do note that there is a plume of potential short duration events visible below our 0.64 year cutoff; these may be true white dwarf microlensing events -- we will examine these further in a future paper.

\subsection{Total Black Hole Population} \label{subsec:total_bh_population}

The results of our search turned up no compelling BH microlensing candidates.  In this section we examine whether this result provides us with any constraints on the BH population of the cluster, first by re-deriving the estimate of the expected number of lensing events, and then by examining the constraints that we have that more compelling candidates were not found.

As noted in the introduction, \citet{Zaris_2020} calculated that a detectable number of microlensing events may be present.  As detailed in Appendix~\ref{app:bh_sampling}, we have improved this calculation in two ways. First, the new microlensing simulations incorporate a realistic number and radial profile of the 2,459 BHs in the \citet{prieto_2025} simulation to get a rate of predicted lensing events over the timespan of our dataset.  Each potential lensing event has a simulated time associated with it, and we calculate the probability of it occurring during our observations by dividing our time span by the simulated time. Second, we can use our completeness shown in Fig.~\ref{fig:magnification_completeness} to evaluate the likelihood of each event being detected in our data (using the "All Events" lines).  We can then add the probabilities of each possible event to get the total number of events detected in our observations.  This results in a total expected number of detected lensing events of just 0.128 during our observations. In this context our lack of any compelling BH microlensing detections is the expected outcome, and our data is consistent with the BH population in the \citet{prieto_2025} simulation.  

An upper limit on the number of microlensing events could provide limits on the BH population in $\omega$~Cen.  Unfortunately, the lack of any strong detection as well as the data shift we see relative to our synthetic lightcurves makes it challenging to place an upper limit on the number of BH microlensing events that may be present in our data.  However, it is clear from
Figs.~\ref{fig:chi_diff} and \ref{fig:outlier_ratios} that many potential synthetic events have larger $\chi^2$ differences and numbers of positive outliers than any lightcurves in our data. In total, 2,829 of the 40,000 simulated lightcurves (7\%) show a higher number of outliers or a greater $\chi^2$ difference than any of the lightcurves we observe. Assuming these synthetics would result in a clear detection in our data, we translate the lack of even a single event like this to an upper limit of $\sim$15 (1/0.07) BH microlensing events with a magnification of $>$0.01 occurred within the time span of our data.

\section*{Acknowledgements}

Work on this paper by Z.F., M.W., A.B., and A.S. was supported by a Hubble Space Telescope archival grant HST-AR-17033, A.S. also acknowledges support from JWST-GO-08322. J.S. acknowledges support from NASA grant 80NSSC21K0628. C.R. acknowledges support from NASA ATP Grant 80NSSC24K0687 and the David and Lucile Packard Foundation. D.V. is grateful to The Scientific and Technological Research Council of T\"urkiye for their support through the grant 123C484. D.V. acknowledges further support from the European Union’s Horizon Europe Research and Innovation Programme under the Marie Skłodowska-Curie Actions (MSCA) COFUND Programme with the grant number 101081645 and The Scientific and Technological Research Council of T\"urkiye with the grant number 123C213. Support for E.G.P. was provided by the NSF Graduate Research Fellowship Program under grant DGE-2234667.

%%%%%%%%%%%%%%%%%%%%%%%%%%%%%%%%%%%%%%%%%%%%%%%%%%
\section*{Data Availability}

The HST data used in this work is publicly available at the MAST archive. In addition, we will make the reduced single epoch photometry available on Zenodo along with the version of reference of the paper once accepted.

%%%%%%%%%%%%%%%%%%%% REFERENCES %%%%%%%%%%%%%%%%%%

\bibliographystyle{mnras}
\bibliography{citations}

%%%%%%%%%%%%%%%%%%%%%%%%%%%%%%%%%%%%%%%%%%%%%%%%%%

%%%%%%%%%%%%%%%%% APPENDICES %%%%%%%%%%%%%%%%%%%%%

\appendix

\section{Microlensing: Simulation Framework}\label{sec:appendix}

\subsection{BH sampling}\label{app:bh_sampling}

To produce realistic lightcurves we have developed a microlensing code that simulates the trajectories of BHs and calculates the resulting magnification of background stars. From the N-body models of \ocen{} computed in \citet{prieto_2025} using CMC \citep{Rodriguez_2022}, we extract the stellar and BH population at 12~Gyr. Specifically, we use the final 12~Gyr timestep of their 500~M$_\odot$ initial seed simulation to simulate our star--BH pairs and create a set of lightcurves characteristic of \ocen{}. An early version of the simulation framework is described in \cite{Zaris_2020}, but we provide a brief overview here. 

First, a single BH is randomly selected from the population. Its mass, distance from the cluster center, and velocity are randomly sampled from the \cite{prieto_2025} simulations and denoted $m_{bh}$, $r_{bh}$, and $v$, respectively. The resulting BH trajectory is assumed to be circular. While actual BH orbits are not circular, this simplification reproduces the cluster's radial density profile in the limit of many simulated BHs. The orientation of the circular BH trajectory is chosen by randomly sampling a unit vector in 3D space, $\hat{q} = (q_x,q_y,q_z)$, which is taken to be orthogonal to the plane of the orbit. Suppose $\hat{a} = (a_x,a_y,a_z)$ and $\hat{b}=(b_x,b_y,b_z)$ are orthogonal unit vectors which are also both orthogonal to $\hat{q}$.  Then a circular trajectory at radius $r_{bh}$ is given by

\begin{equation}\label{eq:bh_pos}
\begin{split}
    &x_{bh} = r_{bh}\cos(\theta)a_x+r_{bh}\sin(\theta)b_x\\
    &y_{bh} = r_{bh}\cos(\theta)a_y+r_{bh}\sin(\theta)b_y\\
    &z_{bh} = r_{bh}\cos(\theta)a_z+r_{bh}\sin(\theta)b_z
\end{split}
\end{equation}

We can write $\theta = 2\pi t/T$ to emphasize the time dependence of the BH motion (note that this $\theta$ is distinct from the definition in equation~(\ref{eq:y_scaling}) - (\ref{eq:lens_eq}), which is the angular position of the deflected light ray in the lens plane), where $T$ is the time required for the BH to complete an orbit ($T=2\pi r_{bh}/v$). Note that since $r_{bh}$ is the distance from the cluster center to the BH, then the position of the BH relative to Earth is $(x_{bh},y_{bh}, z_{bh}+D_c) = \vec{x}_{bh}+D_c\hat{z}$. 

Stars are taken to be stationary in our code. In particular, the motion of stars is incorporated within the BH trajectory.  We replace $T$ with $T'=2\pi r_{bh}/v_{rel}$ where $v_{rel}$ is the relative velocity between the BH and star (randomly sampled using the speed of both objects). The BH is essentially evolved along its trajectory using the relative velocity. This is justified since the relative motion between the lens and light source determines the shape of a lightcurve.

Each simulation run uses a single BH and a single star.  However, the vast majority of stars are located far from the BH orbit and will not result in detectable lensing events. Therefore, we implement a cut criterion for stars so that computational resources are not used to simulate non-events. Stars are generated one at a time by randomly sampling their radii and velocities (denoted $r_s$ and $v_s$, respectively) from the \cite{prieto_2025} stellar distribution. First, the position of the star is projected onto the plane of the BH orbit by finding the point at which the line of sight from Earth to the star intersects the BH plane. Then the minimum angular separation between this point and the BH orbit is approximated. If this angle is sufficiently large, then the star is rejected, otherwise, it is accepted. Mathematically, the line of sight from Earth to the star, whose position relative to Earth is $(x_s,y_s,z_s+D_c) = \vec{x}_s+D_c\hat{z}$ is given by the vector

\begin{equation}
\begin{split}
    \vec{l}_s = \lambda(x_s\hat{x}+y_s\hat{y}+(z_s+D_c)\hat{z})
\end{split}
\end{equation}

The value of $\lambda$ at which the line intersects the plane is then given by

\begin{equation}
\begin{split}
    \lambda_{int} = \frac{D_c\hat{z}\cdot\hat{q}}{(x_s\hat{x}+y_s\hat{y}+(z_s+D_c)\hat{z})\cdot\hat{q}}
\end{split}
\end{equation}

So the intersection point, with Earth at the origin, is $\vec{x}_{int} = (\lambda_{int}x_s,\lambda_{int}y_s,\lambda_{int}(z_s+D_c))$. To approximate the minimum angular distance from $\vec{x}_{int}$ to the BH orbit, we use the distance between $\vec{x}_{int}$ and the orbit along the line connecting the center of the cluster to $\vec{x}_{int}$. We define 

\begin{equation}
\begin{split}
    \vec{\xi}_{bh} = \frac{r_{bh}}{|\vec{x}_{int}-D_c\hat{z}|}(\vec{x}_{int}-D_c\hat{z})
\end{split}
\end{equation}

The angular distance is then approximately 

\begin{equation}
\begin{split}
    \psi = \frac{|(\vec{x}_{int}-\vec{\xi}_{bh})\times\hat{z}|}{D_c}
\end{split}
\end{equation}

\begin{figure*}
    \centering
    \includegraphics[scale=0.53]{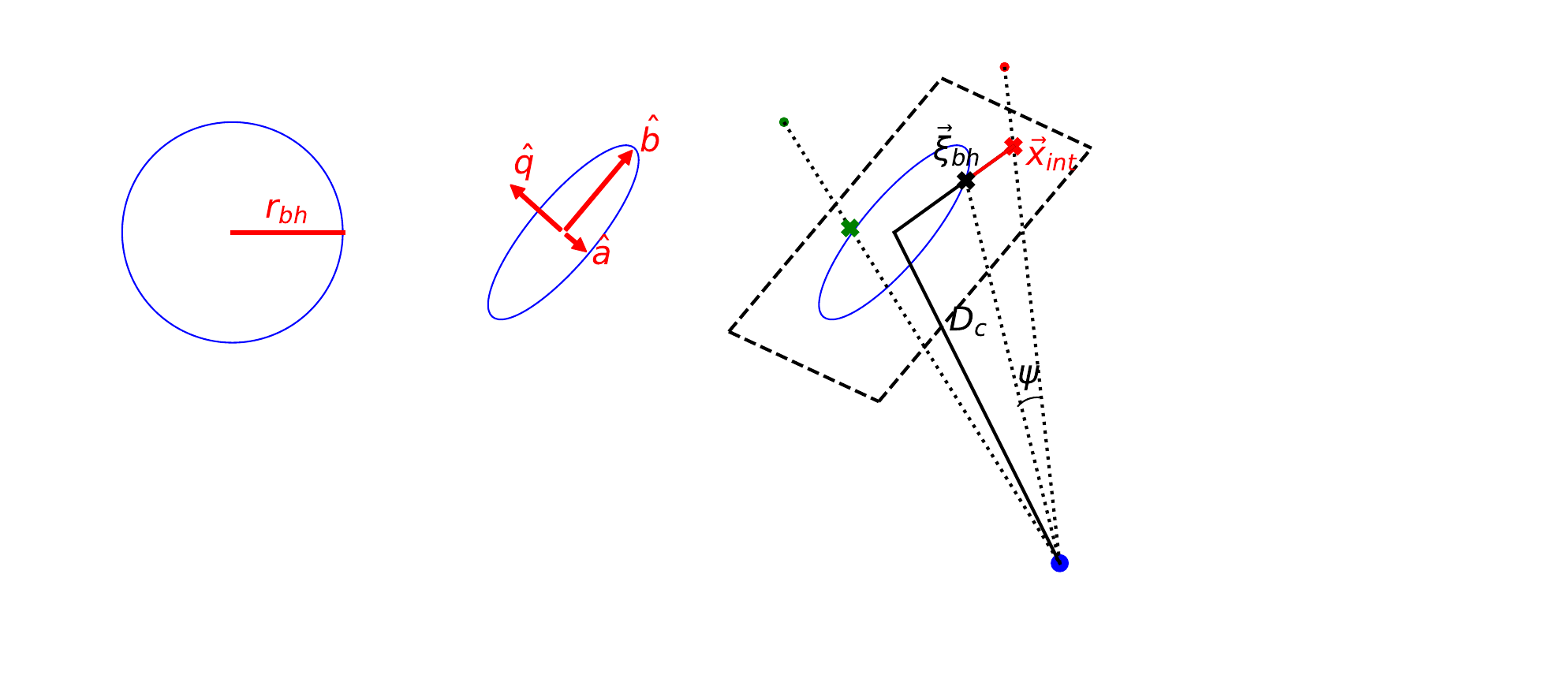}
    ~ 
    \caption{The radius of the BH orbit about the center of the cluster is randomly sampled according to simulations from \citet{prieto_2025}.  Next, the orientation of the orbit is randomized, resulting in the orbit appearing elliptical from an observer on earth.  The semi-major axis of this ellipse is always equal to the orbital radius $|b| = r_{bh}$.  Stars are then generated randomly, and most are located too far from the ellipse to be lensed.  The process for evaluating stars is as follows: the star is projected (along the line of sight from Earth to the star) onto the plane which contains the BH orbit and this point is denoted $\vec{x}_{int}$.  Then the angular distance $\psi$ between this point and the BH orbit is roughly approximated.  We find the intersection of the line connecting the center of the orbit and $\vec{x}_{int}$ with the BH orbit and denote this point $\vec{\xi}_{bh}$. Then $\psi$ is approximated as the angular separation between $\vec{x}_{int}$ and $\vec{\xi}_{bh}$, as viewed from Earth.}
    \label{fig:cooling_setup}
\end{figure*}

After testing various cutoffs, we have chosen to reject stars with $\psi>{2\times}10^{-8}=\epsilon$. Setting the value of $\epsilon$ too small would result in stars being cut which would have been lensed. Setting the value too high is inefficient as it leads to the simulation of many nonevents. We illustrate the star sampling process in the last illustration in Fig.~\ref{fig:cooling_setup}.

Next, the BH position is initialized by setting $\theta=0$ in equation~(\ref{eq:bh_pos}) and $v_{rel}$ is randomly sampled according to the individual star and BH velocities. The BH position is evolved by setting $\theta= 2\pi t/T'$ in equation~(\ref{eq:bh_pos}). The variable timestep used in the simulations is parameterized by $\Delta t = t_c/v_{rel}$, where $t_c$ is a constant which is sufficiently small to capture even relatively brief lensing events. 

At each timestep, the code calculates the magnification $\mu$ of the star caused by the BH, given by

\begin{equation}
\begin{split}
    \mu = \frac{u^2+2}{u\sqrt{u^2+4}}
\end{split}
\end{equation}

Here, $u$ is the impact parameter. Denoting the angular separation of the BH and star as $\alpha$, the impact parameter is $u=\alpha/\theta_E$, where $\theta_E$ is the angular Einstein radius expressed as

\begin{equation}
\begin{split}
    \theta_E = \sqrt{\frac{4Gm_{bh}(\vec{D}_s'-|\vec{x}_{bh}+D_c\hat{z}|)}{c^2|\vec{D}_s'||\vec{x}_{bh}+D_c\hat{z}|}},
\end{split}
\end{equation}
where we have defined
\begin{equation}
\begin{split}
    \vec{D}_s'= \frac{(\vec{x}_s+D_c\hat{z})\cdot (\vec{x}_{bh}+D_c\hat{z})}{|\vec{x}_{bh}+D_c\hat{z}|}
\end{split}
\end{equation}

Therefore, the impact parameter is explicitly calculated as

\begin{equation}
\begin{split}
    u = \frac{|\vec{x}_s+D_c\hat{z} - (\vec{x}_{bh}+D_c\hat{z})(\frac{|\vec{x}_s+D_c\hat{z}|}{|\vec{x}_{bh}+D_c\hat{z}|})|}{|\vec{x}_s+D_c\hat{z}|\theta_E}
\end{split}
\end{equation}

For almost the entirety of the trajectory this magnification is negligible, but it may reach the specified threshold value of $\mu_{th}$ when the BH approaches the star. If it is found that $\mu\geq \mu_{th}$ at any point along the BH trajectory then a lensing event is deemed to occur. In order to accelerate our code we use an adaptive timestep.  Specifically, in $\Delta t = t_c/v_{rel}$, we decrease $t_c$ during lensing events and increase $t_c$ when lensing is not occurring.  Furthermore, we can approximate the value of $\theta$ at which the lensing occurs, so we don't need to integrate the entire circular trajectory.  Instead, we can simulate a small portion of the trajectory where the largest impact parameter occurs.

\subsection{Neural Network Training Details}\label{app:nn_training}

Our neural network is a fully connected, feedforward multilayer perceptron with no skip or residual connections and no normalization layers (e.g., batch or layer normalization). We performed a hyperparameter grid search over five hyperparameters in our neural network: the number of hidden layers (8, 16, 32), the number of neurons per hidden layer (32, 64, 128), the learning rate (0.01, 0.001, 0.0001), the weight decay (0.0001, 0.00001, as implemented in PyTorch \citep{pytorch}), and the neuron dropout rate (0, 0.2); a total of 108 hyperparameter combinations. This grid search was implemented with a custom training loop rather than an external hyperparameter optimization library. Our final hyperparameters are 16 hidden layers, 128 neurons per hidden layer, a learning rate of 0.001, a weight decay of 0.00001, and a dropout rate of 0. We use ReLU as the activation function in the hidden layers and train for 30 epochs (confirmed as a sufficient number of epochs by convergence of validation loss) using the Adam optimizer \citep{kingma2017adammethodstochasticoptimization} with batch sizes of 2048. Network weights are initialized using PyTorch's default initialization scheme. Since our task is a regression problem in which the network predicts a probability, the output layer consists of a single neuron, and we train using a binary cross-entropy loss. Training was performed on a single NVIDIA RTX 2080 Ti GPU. All models were implemented in Python \texttt{3.12.9} using PyTorch \texttt{2.10.0} \citep{pytorch}.

We used 23 input features in the neural network. A subset of these input features can be missing; this stems from scenarios where a lensing fit gets a very low magnification. Since a model fit with very low magnification (<$1.01$) doesn't have a meaningful duration, we are unable to calculate e.g., the number of bright outliers within a fit ($n_{bright-outlier}$ in Table~\ref{table:nn_features}). However, it is interesting and helpful to keep these features that are missing for the neural network, since it allows the neural network to learn a representation for lightcurves that have flat model fits. We therefore use a missing indicator for all values that are able to be missing; the original value is then set to $-10$, a value well outside the range of acceptable values. These missing indicators are then used as additional input features to the neural network. A description of each of these features can be found in Table~\ref{table:nn_features}.

We measured the correlation between the features and the classification of the synthetic lightcurves (i.e., positive or negative events) using the \textit{mutual information score}, a statistic that can capture correlation in non-linear relationships. All features had low mutual information scores, the highest mutual information score being $\chi^2_{l-m}$ with a value of 0.118. The second and third highest mutual information scores were the initial condition tier ($0.08$) and $\chi^2_{c-m}$ ($0.073$), respectively.

\begin{table*}
	\centering
	\caption{The input features to the neural network.}
	\label{table:nn_features}
	\begin{tabular}{p{0.25\textwidth} p{0.65\textwidth}}
		\hline
		  Feature & Description\\
		\hline
		$\chi^2_{red.,model}$ & The reduced chi square value of the model fit. \\
        $\chi^2_{red.,constant}$ & The reduced chi square value of the constant line fit (i.e., zero slope). \\
        $\chi^2_{red.,line}$ & The reduced chi square value of the line fit. \\
        $\chi^2_{c-m}$ & $\chi^2_{red.,constant} - \chi^2_{red.,model}$ \\
        $\chi^2_{l-m}$ & $\chi^2_{red.,line} - \chi^2_{red.,model}$ \\
        $\chi^2_{l-c}$ & $\chi^2_{red.,line} - \chi^2_{red.,constant}$ \\
        $n_{bright-fit}$ & The number of $>2\sigma_G$ bright outliers within the duration of the fit lensing event. \\
        $n_{bright-out}$ & The number of $>2\sigma_G$ bright outliers outside of the duration of the fit lensing event. \\
        $n_{dim-fit}$ & The number of $>2\sigma_G$ dim outliers within the duration of the fit lensing event. \\
        $n_{in-fit}$ & The total number of observations within the fit. \\
        $R_{outlier}$ & The outlier ratio, calculated by $\frac{n_{bright-fit}}{n_{bright-out} + n_{dim-fit}}$. Range of [0, $n_{bright-fit}$].  \\
        $R_{epoch}$ & The epoch ratio, calculated by $\frac{n_{bright-fit} - n_{dim-fit}}{n_{in-fit}}$. Range of [-1, 1]. \\
        $n_{bright-fit}$ Missing Indicator & A binary value that indicates whether $n_{bright-fit}$ is missing. \\
        $n_{bright-out}$ Missing Indicator & A binary value that indicates whether $n_{bright-out}$ is missing. \\
        $n_{dim-fit}$ Missing Indicator & A binary value that indicates whether $n_{dim-fit}$ is missing. \\
        $n_{in-fit}$ Missing Indicator & A binary value that indicates whether $n_{in-fit}$ is missing. \\
        $R_{outlier}$ Missing Indicator & A binary value that indicates whether $R_{outlier}$ is missing. \\
        $R_{epoch}$ Missing Indicator & A binary value that indicates whether $R_{epoch}$ is missing. \\
        Slope & The slope fit by the line fit. \\
        Magnification & The magnification fit by the model fit. \\
        Transverse Velocity & The transverse velocity fit by the model fit. \\
        Number of bright exposure clusters & This feature counts the total number of bright exposure clusters; as defined in Section \ref{subsubsec:initial_params}, a bright exposure cluster is $\geq 3$ bright exposures within a 0.2 year range. \\ 
        Initial Condition Tier & The initial condition tier used, as defined in Section \ref{subsubsec:initial_params}. \\
		\hline
	\end{tabular}
\end{table*}

%%%%%%%%%%%%%%%%%%%%%%%%%%%%%%%%%%%%%%%%%%%%%%%%%%

\bsp
\label{lastpage}
\end{document}